\documentclass[12pt]{article}
\usepackage{graphicx,verbatim,array,multicol,courier}
\usepackage{amsmath,amssymb,amsthm,mathrsfs,mathdots,mathtools}
\usepackage[pdftex,bookmarks,colorlinks]{hyperref}
\usepackage{fullpage}
\usepackage{topcapt}
\usepackage{color}
\usepackage{bbm}
\usepackage{dsfont}
\usepackage[section]{placeins}
\usepackage[labelformat=simple]{subfig}
\usepackage{soul}
\usepackage{booktabs}
\usepackage{epstopdf}
\usepackage{natbib}
\usepackage{enumitem}
\usepackage{booktabs, arydshln}
\usepackage{lscape}
\theoremstyle{definition}
\newtheorem{defi}{Definition}[section]

\theoremstyle{definition}
\newtheorem{cond}{Condition}[section]

\theoremstyle{plain}
\newtheorem{Theo}[defi]{Theorem}

\newtheorem{Coro}[defi]{Corollary}

\theoremstyle{definition}

\newcommand{\bfgamma}{{\boldsymbol{\gamma}}}

\newcommand{\bfSigma}{{\boldsymbol{\Sigma}}}

\newcommand{\bfxitilde}{{\smash{\widetilde{\boldsymbol{\xi}}\vphantom{()}}}}
\newcommand{\bfxihat}{{\smash{\widehat{\boldsymbol{\xi}}\vphantom{()}}}}
\newcommand{\bfqhat}{{\smash{\widehat{\boldsymbol{q}}\vphantom{()}}}}
\newcommand{\bfVhat}{{\smash{\widehat{\boldsymbol{V}}\vphantom{()}}}}
\newcommand{\bfSigmahat}{{\smash{\widehat{\boldsymbol{\Sigma}}\vphantom{()}}}}
\newcommand{\bfbhat}{{\smash{\widehat{\boldsymbol{b}}\vphantom{()}}}}

\newcommand{\bfb}{{\boldsymbol{b}}}

\newcommand{\bfC}{{\boldsymbol{C}}}

\newcommand{\bfR}{{\boldsymbol{R}}}
\newcommand{\bfu}{{\boldsymbol{u}}}
\newcommand{\bfU}{{\boldsymbol{U}}}

\newcommand{\bfV}{{\boldsymbol{V}}}

\newcommand{\bfX}{{\boldsymbol{X}}}

\newcommand{\bfzero}{{\boldsymbol{0}}}

\newcommand{\diff}{\mathrm{d}}

\newcommand{\surv}{\overline{F}}

\newcommand{\tod}{\stackrel{\mathrm{d}}{\longrightarrow}}

\newcommand{\ind}{\mathbbm{1}}
\newcommand{\EE}{\mathbb E}
\newcommand{\PP}{\mathbb P}
\newcommand{\RR}{\mathbb R}

\DeclareMathOperator*{\argmin}{arg\,min}

\DeclarePairedDelimiter\floor{\lfloor}{\rfloor}

\makeatletter
\def\adl@drawiv#1#2#3{%
        \hskip.5\tabcolsep
        \xleaders#3{#2.5\@tempdimb #1{1}#2.5\@tempdimb}%
                #2\z@ plus1fil minus1fil\relax
        \hskip.5\tabcolsep}
\newcommand{\cdashlinelr}[1]{%
  \noalign{\vskip\aboverulesep
           \global\let\@dashdrawstore\adl@draw
           \global\let\adl@draw\adl@drawiv}
  \cdashline{#1}
  \noalign{\global\let\adl@draw\@dashdrawstore
           \vskip\belowrulesep}}
\makeatother

\newlist{inparaenum}{enumerate}{2}
\setlist[inparaenum,1]{label=(\roman*)}
\setlist[inparaenum,2]{label=(\roman{inparaenumi}\emph{\alph*})}

\definecolor{navy}{rgb}{0,0,0.502}
\definecolor{brown}{rgb}{0.59, 0.29, 0.0}

\hypersetup{colorlinks,%
citecolor=blue,%
filecolor=green,%
linkcolor=red,%
urlcolor=magenta,%
}
\title{Joint inference on extreme expectiles for multivariate heavy-tailed distributions}
\author{Simone A. Padoan\footnote{Simone A. Padoan works at the
Department of Decision Sciences,
Bocconi University of Milan, via Roentgen 1, 20136 Milano, Italy,
E-mail: {\tt simone.padoan@unibocconi.it}.} \, and Gilles Stupfler\footnote{Gilles Stupfler works at Univ Rennes, Ensai, CNRS, CREST - UMR 9194, F-35000 Rennes, France,
E-mail: {\tt gilles.stupfler@ensai.fr}.}}

\begin{document}
\maketitle
\begin{abstract}
The notion of expectiles, originally introduced in the context of testing for homoscedasticity and conditional symmetry of the error distribution in linear regression, induces a law-invariant, coherent and elicitable risk measure that has received a significant amount of attention in actuarial and financial risk management contexts. 
A number of recent papers have 
focused on the behaviour and estimation of extreme expectile-based risk measures and their potential for risk management. Joint inference of several extreme expectiles has however been left untouched; in fact, even 
the inference of a marginal extreme expectile turns out to be a difficult problem in finite samples. We investigate the simultaneous estimation of several extreme marginal expectiles of a random vector with heavy-tailed marginal distributions. This is done in a general extremal dependence model where the emphasis is on pairwise dependence between the margins. We use our results to derive accurate confidence regions for extreme expectiles, as well as a test for the equality of several extreme expectiles. Our methods are showcased in a finite-sample simulation study and on real financial data.
\end{abstract}
%
\section{Introduction and background}\label{sec:intro}
%
Expectiles, introduced by~\cite{newpow1987}, induce risk measures which have recently gained substantial traction in the risk management context. Expectiles of an integrable random variable $X$ are obtained as minimisers of asymmetrically squared deviations in the following sense: 
\begin{equation}
\label{eqn:expectile}
\xi_{\tau} = \argmin_{\theta\in \RR} \EE(\eta_{\tau}(X-\theta)-\eta_{\tau}(X)),
\end{equation}
where $\eta_{\tau}(u)=|\tau-\ind\{ u\leq 0 \}| u^2$ is the so-called expectile check function and $\ind\{ \cdot \}$ the indicator function. Expectiles can be seen as $L^2-$analogues of quantiles, which can be obtained by minimising asymmetrically weighted mean absolute deviations~\citep{koebas1978}:
\[
q_{\tau} \in \argmin_{q\in \RR} \EE(\rho_{\tau}(X-q)-\rho_{\tau}(X)),
\]
where $\rho_{\tau}(u)=|\tau-\ind\{ u\leq 0 \}| |u|$ is the quantile check function. Unlike the $\tau$th quantile, the $\tau$th expectile is always uniquely defined by its convex optimisation problem, and satisfies
\begin{equation}
\label{eqn:gainloss}
\tau=\EE[|X-\xi_{\tau}|\ind\{ X\leq \xi_{\tau}\} ]/\EE|X-\xi_{\tau}|. 
\end{equation}
In particular, expectiles 
depend on tail realisations of the loss variable as well as their probability. The advantages of the expectile include that it is the only risk measure, apart from the simple expectation, that defines a law-invariant, coherent~\citep{artdelebehea1999} and elicitable~\citep{gne2011} risk measure, see~\cite{belklamulgia2014} and~\cite{zie2016}. It follows from the elicitability property that expectiles benefit from the existence of a natural backtesting methodology. Quantiles, by contrast, are elicitable, but are often criticised for not being a coherent risk measure, and for missing out on important information about the tail of the underlying distribution since they only depend on the frequency of tail events. Meanwhile, 
the popular quantile-based Expected Shortfall is coherent, takes into account the actual values of the risk variable on the tail event, but is not elicitable.
Formula~\eqref{eqn:gainloss} links expectiles to the notion of gain-loss ratio, which is a popular performance measure in portfolio management and is well-known in the literature on no good deal valuation in incomplete markets~\citep[see][and references therein]{beldib2017}. Further investigations carried out by~\cite{ehmgnejorkru2016} and~\cite{beldib2017}, among others, suggest that expectiles define perfectly sensible alternatives to the quantile and Expected Shortfall.

Although expectile estimation dates back to~\cite{newpow1987} in the context of linear regression, it has been the subject of renewed interest in a large range of models, see for example~\cite{sobkne2012} and references therein as well as~\cite{holkla2016} and~\cite{krazah2017}, for the estimation of central expectiles of fixed order $\tau$ staying away from the tails of the underlying distribution. 
Meanwhile, probabilistic aspects of extreme expectiles, with $\tau\uparrow 1$, have been examined by~\cite{belklamulgia2014} and~\cite{beldib2017}. Inference on extreme expectiles has been considered even more recently in~\cite{daogirstu2018,daogirstu2019,daogirstu2020}. These results are limited to inference about extreme expectiles of a single sample of data; in other words, they do not make it possible to construct joint confidence regions for several extreme expectiles from different variables of interest. This is a substantial restriction in actuarial and financial applications, where practitioners are interested in evaluating the asymptotic dependence existing within several risk variables, stock prices or stock indices, and in carrying out precise joint inference about the extremes of these risk variables. Such questions are for instance considered in~\cite{jonpurzit2006} with nonparametric testing of equality of distortion risk measures in an actuarial context, in~\cite{strverwol2008} in the detection of tail asymmetries,~\cite{zho2010} and~\cite{maimitrus2015} for the construction of diversified financial portfolios, 
and in~\cite{hurlauquasme2017} 
as a way to directly compare risk measures between assets. Besides, an inspection of the results in~\cite{daogirstu2018} shows that, in the univariate case, standard plug-in asymptotic confidence intervals obtained from the asymptotic normality of the estimators behave in fact more often than not quite poorly in finite samples. In particular, the Gaussian QQ-plots in Appendix~A.2 of~\cite{daogirstu2018} show that, despite the fact that the Gaussian distribution will in many cases be a reasonable model for the uncertainty of extreme expectile estimators, the sample variance of the estimators can be a long way off the variance obtained {\it via} a naive use of the theoretical Gaussian approximation. These two issues constitute a serious gap that should be addressed if expectiles are to be used widely in risk management.

This paper contributes to filling that gap as follows. In a general framework of multivariate distributions with marginal heavy tails and extremal dependence between margins, and given independent and identically distributed (i.i.d.)~data, we start by rigorously investigating the joint asymptotic normality of intermediate tail expectiles of the margins. The order of expectiles is such that $\tau=\tau_n \uparrow 1$ with $n(1-\tau_n) \to \infty$ as $n\to\infty$, where $n$ denotes sample size. 
Let us highlight 
that the theoretical properties of the methods we shall consider, called the Least Asymmetrically Weighted Squares (LAWS) estimators and Quantile-Based (QB) estimators, had been analysed only for the estimation of a single extreme expectile. Our emphasis here is on describing the asymptotic dependence structure of our estimators using the concept of tail copula introduced and studied in~\cite{schsta2006}.
Our results are then used to tackle the important question of joint inference about tail expectiles from two distinct angles. First, we exploit our joint Gaussian asymptotics of tail expectile estimators to construct asymptotic joint confidence regions for tail expectiles. This is done by, on the one hand, designing specific finite-sample corrections for the standard plug-in asymptotic variance estimators of each expectile estimator to obtain accurate representations of marginal uncertainty. On the other hand, we construct an appropriate nonparametric estimator of the tail dependence between two such estimators pertaining to different marginals. This results in an estimate of the covariance matrix of our set of expectile estimators, used to build Gaussian confidence regions for the vector of expectiles of interest and resulting in a procedure that is computationally very fast and avoids having to resort to bootstrapping. Second, we tackle the important problem of testing whether tail expectiles across marginals are equal. We do so by adapting the classical likelihood ratio test of equal means in a Gaussian random vector. The deviance statistic in this testing procedure prominently features our covariance matrix estimators that will be used to construct accurate confidence regions. 

The outline of the paper is the following. Section~\ref{sec:main} explains in detail our statistical context and contains the main theoretical results of the paper on joint intermediate and extreme expectile estimation. Section~\ref{sec:CI} explores the implications of our results on joint inference about tail expectiles. The finite-sample performance of the methods is examined on simulated data sets in Section~\ref{sec:simulations} and on financial exchange rates data in Section~\ref{sec:real_data}. The methods and data considered in this article have been incorporated into the {\color{magenta}{\tt R}} package {\color{magenta}{\tt ExtremeRisks}}, freely available on CRAN. 
The 
Appendix gives further finite-sample results.

%
%
\section{Joint estimation of multiple extreme expectiles}\label{sec:main}
%

Let $(\bfX_i, 1\leq i\leq n)$, with $\bfX_i=(X_{i,j},1\leq j\leq d)$, be i.i.d.~copies of a $d$-dimensional random vector $\bfX=(X_1,\ldots,X_d)$, 
with marginal distributions $F_{j}$, associated survival functions $\surv_j=1-F_j$, and tail quantile functions $U_j(s)=\inf\{ x\in \RR \, | \, F_j(x)\geq 1-s^{-1}\}$, for $s>1$. The realisations of $X_j$ may for example be seen as the negatives of generic financial positions, so that large positive values of $X_j$ represent extreme losses associated to one specific position, or as losses incurred by an insurance company in distinct lines of business.

We focus on the joint estimation of extreme expectiles of $X_1,\ldots,X_d$. 
We work with heavy-tailed distributions, representing the tail structure of many financial and actuarial data examples fairly well, 
see {\it e.g.}~p.9 of~\cite{embklumik1997}. Mathematically, we assume
\[
\forall j\in \{1,\ldots,d\}, \ \forall x>0, \ \lim_{s\to\infty}\frac{\surv_j(sx)}{\surv_j(s)} = x^{-1/\gamma_j} \ \mbox{ or equivalently } \ \lim_{s\to\infty}\frac{U_j(sx)}{U_j(s)} = x^{\gamma_j}.
\]
The tail indices $\gamma_j>0$ specify marginal tail heaviness. With condition $\EE|\min(X_j,0)|<\infty$, the assumption $0<\gamma_j<1$ ensures that the first moment of $X_j$ exists and thus expectiles of the $X_j$ are well-defined. These two conditions will be part of our assumptions throughout.

More precisely, our overarching focus in the present paper is to establish the joint asymptotic distribution of tail expectile estimators of level $\tau$ close to 1. Specifically, according to~\eqref{eqn:expectile}, the expectile for the $j$th marginal distribution $F_{j}$ is defined as
\begin{equation}
\label{eqn:expectile_j}
\xi_{\tau,j}= \argmin_{\theta\in\RR} \EE \left( \eta_{\tau}(X_j-\theta) - \eta_{\tau}(X_j)\right),
\end{equation}
where $\eta_{\tau}$
is the expectile check function defined below Equation~\eqref{eqn:expectile}. We consider hereafter the problem of the joint inference of $(\xi_{\tau,1},\ldots,\xi_{\tau,d})$, where the level $\tau$ is such that $\tau=1-p$ for a small value of $p=p_n$. Two cases are considered, when $p$ is (much) larger and smaller than $1/n$, with $n$ large: these are respectively the {\it intermediate case}, when nonparametric estimation methods can be used, and the properly {\it extreme case} when extrapolation methods whose rationale is rooted in the heavy-tailed assumption have to be developed. To carry out joint inference about estimators of extreme expectiles, we model here the extremal dependence structure between any two components of $\bfX$ in the form of a tail copula. This translates into the following general assumption that we shall work with throughout.
\begin{cond}\label{cond:multivariate}
For every $1\leq j\leq d$, let $F_j$ and $U_j$ be the distribution function and tail quantile function associated to $X_j$. Assume that the $F_j$ are continuous and: 
\begin{inparaenum}
\item \label{cond:enu:reg_varying} $U_j$ is regularly varying with index $\gamma_j$: $U_j(sx)/U_j(s) \to x^{\gamma_j}$ as $s\to\infty$, for any $x>0$. 
%
%
\item \label{cond:enu:tail_copula} For any $(j,\ell)$ with $j\neq \ell$, there is a function $R_{j,\ell}$ on $[0,\infty]^2 \setminus \{  (\infty,\infty)\}$ such that
\[
\forall (x_j,x_{\ell})\in [0,\infty]^2 \setminus \{  (\infty,\infty)\}, \ \lim_{s\to \infty} s \, \PP\left( \surv_j(X_j)\leq \frac{x_j}{s}, \ \surv_\ell(X_\ell)\leq \frac{x_\ell}{s} \right) = R_{j,\ell}(x_j,x_\ell).
\]
\end{inparaenum}
\end{cond}
Condition~\ref{cond:multivariate}\ref{cond:enu:tail_copula} formalises the existence of a limiting dependence structure in the upper tail of any two components $X_j$ and $X_{\ell}$, given by the {\it tail copula} $R_{j,\ell}$~\citep{schsta2006}. It is a weak assumption since it is satisfied by any $\bfX$ in the maximum domain of attraction of a multivariate extreme value distribution~\citep{haafer2006}.

%
%
\subsection{At the intermediate level}\label{sec:multiple_inter}
%
Let $0<\tau_n<1$ satisfy $\tau_n\to 1$ and $n(1-\tau_n)\to \infty$ as $n\to\infty$. We focus on estimating tail expectiles of the $X_j$ at level $\tau_n$. We consider two methods: the nonparametric empirical counterpart of~\eqref{eqn:expectile_j}, called the {\it Least Asymmetrically Weighted Squares} (LAWS) estimator and a semiparametric {\it Quantile-Based} (QB) estimator built on our heavy-tailed assumption. 
\vskip1ex
\noindent
{\bf Nonparametric estimator via asymmetric least squares \hskip1ex} We first consider estimating the expectile $\xi_{\tau_n,j}$ of the marginal distribution $F_j$ by its empirical estimator
\begin{equation*}
\widetilde{\xi}_{\tau_n,j} = \argmin_{\theta\in \RR} \sum_{i=1}^n \eta_{\tau_n}(X_{i,j}-\theta).
\end{equation*}
This LAWS estimator can be computed with iteratively reweighted least squares, or with standard minimisation routines such as {\tt uniroot} in {\tt R}. Theorem 2 in~\cite{daogirstu2018} shows that the empirical estimator $\widetilde{\xi}_{\tau_n,j}$ is consistent and $\sqrt{n(1-\tau_n)}-$asymptotically normal; this result is limited to the marginal estimation of an intermediate expectile. Our first main result provides the joint asymptotic normality of the estimators $\widetilde{\xi}_{\tau_n,j}$, for $1\leq j\leq d$.
\begin{Theo}\label{theo:asydirinter}
Assume that Condition~\ref{cond:multivariate} is satisfied. Assume further that there is $\delta>0$ such that $\EE|\min(X_j,0)|^{2+\delta}<\infty$ and that $0<\gamma_j<1/2$ for any $1\leq j\leq d$. Let $\tau_n\uparrow 1$ be such that $n(1-\tau_n)\to\infty$ as $n\to\infty$. Then we have 
\[
\sqrt{n(1-\tau_n)} \left( \frac{\widetilde{\xi}_{\tau_n,j}}{\xi_{\tau_n,j}} - 1 \right)_{1\leq j\leq d} \tod \mathcal{N}_d\left(\bfzero_d, \bfV^{\mathrm{LAWS}}(\bfgamma,\bfR) \right). 
\]
The covariance matrix $\bfV^{\mathrm{LAWS}}(\bfgamma,\bfR)$ has entries
\[
\bfV_{j,\ell}^{\mathrm{LAWS}}(\bfgamma,\bfR) =\begin{cases} \dfrac{2\gamma_j^3}{1-2\gamma_j} & \mbox{if } j=\ell, \\[15pt] \gamma_j \gamma_\ell \displaystyle\iint_{[1,\infty)^2} R_{j,\ell}\left((\gamma_j^{-1}-1) x_j^{-1/\gamma_j},(\gamma_\ell^{-1}-1) x_{\ell}^{-1/\gamma_\ell} \right) \diff x_j \, \diff x_{\ell} & \mbox{if } j<\ell. \end{cases}
\]
\end{Theo}
%
To understand the above joint asymptotic distribution further, consider the case $\gamma_j=\gamma$ for any $j\in \{1,\ldots,d\}$, when the $X_j$ have equivalent tails. By 1-homogeneity of the tail copula~\citep[see][Theorem~1(ii)]{schsta2006}, we have
\[
\bfV_{j,\ell}^{\mathrm{LAWS}}(\bfgamma,\bfR) =\begin{cases} \dfrac{2\gamma^3}{1-2\gamma} & \mbox{if } j=\ell, \\[10pt] \gamma (1-\gamma) \displaystyle\iint_{[1,\infty)^2} R_{j,\ell}\left(x_j^{-1/\gamma}, x_{\ell}^{-1/\gamma} \right) \diff x_j \, \diff x_{\ell} & \mbox{if } j<\ell. \end{cases}
\]
The variance term on the diagonal of this matrix is indeed equal to the asymptotic variance derived in~\citet[][Theorem~2]{daogirstu2018} in the univariate case. The covariance terms off the diagonal can be rewritten in terms of the asymptotic correlation of two estimators as
\begin{align*}
\bfV_{j,\ell}^{\mathrm{LAWS}}(\bfgamma,\bfR) &= \dfrac{2\gamma^3}{1-2\gamma} \times \bfC_{j,\ell}^{\mathrm{LAWS}}(\bfgamma,\bfR) \\[5pt]
\mbox{with } \bfC_{j,\ell}^{\mathrm{LAWS}}(\bfgamma,\bfR) &= \frac{1}{2} \times \frac{(1-\gamma)(1-2\gamma)}{\gamma^2} \displaystyle\iint_{[1,\infty)^2} R_{j,\ell}\left(x_j^{-1/\gamma}, x_{\ell}^{-1/\gamma} \right) \diff x_j \, \diff x_{\ell}.
\end{align*}
The expression of the correlation structure $\bfC^{\mathrm{LAWS}}(\bfgamma,\bfR)$ is similar to the one representing the contribution of temporal dependence in the variance of the intermediate marginal LAWS expectile estimator in a stationary time series, see~\citet[][Theorem~3.1]{padstu2020}. 
%
%
\vskip1ex
\noindent
{\bf Semiparametric estimator via a quantile-based procedure \hskip1ex} An alternative estimator is provided by the asymptotic proportionality relationship between expectile and quantile:
\begin{equation}
\label{eqn:expectiles_connection}
\lim_{\tau\uparrow 1} \frac{\xi_{\tau,j}}{q_{\tau,j}} = (\gamma_j^{-1}-1)^{-\gamma_j},
\end{equation}
where $q_{\tau,j}$ is the quantile function of the $j$th marginal. This was first noted by~\cite{belklamulgia2014}. This connection suggests the class of QB estimators
\[
\widehat{\xi}_{\tau_n,j} = (\widehat{\gamma}^{-1}_{\tau_n,j}-1)^{-\widehat{\gamma}_{\tau_n,j}} \widehat{q}_{\tau_n,j}
\]
where for each $j\in \{1,\ldots,d\}$, $\widehat{q}_{\tau_n,j}$ and $\widehat{\gamma}_{\tau_n,j}$ are consistent estimators of $q_{\tau_n,j}$ and $\gamma_j$. Throughout, the estimator $\widehat{q}_{\tau_n,j}$ is taken to be $\widehat{q}_{\tau_n,j}=X_{n-\floor*{n(1-\tau_n)},n,j}$, where $\floor*{\cdot}$ is the floor function and $X_{1,n,j}\leq\cdots\leq X_{n,n,j}$ denote the ascending order statistics of the sample $(X_{1,j},\ldots X_{n,j})$. There has been a wealth of research on the estimation of the tail index $\gamma_j$; we refer to Chapter~4 in~\cite{beigoesegteu2004} and Chapter~3 of~\cite{haafer2006}. We work below with the Hill estimator \citep{hil1975} with effective sample size $k=\floor*{n(1-\tau_n)}$:
\[
\widehat{\gamma}_{\tau_n,j}=\frac{1}{\floor*{n(1-\tau_n)}}\sum_{i=1}^{\floor*{n(1-\tau_n)}}\log\left(\frac{X_{n-i+1,n,j}}{X_{n-\floor*{n(1-\tau_n)},n,j}}\right).
\]
This estimator is in fact the maximum likelihood estimator in the purely Pareto model and is known to be optimal, in terms of rate of convergence, when the distribution function $F_j$ belongs to the wide Hall-Welsh class of models~\citep{halwel1985}, that is 
\begin{equation}
\label{eqn:halwel} 
\surv_j(x)=x^{-1/\gamma_j} \left( a_j+b_j x^{\rho_j/\gamma_j}+\operatorname{o}(x^{\rho_j/\gamma_j}) \right) \mbox{ as } x\to\infty,
\end{equation}
where $a_j>0$, $b_j\neq 0$ and $\rho_j<0$. See~\cite{dre1998}. 

The asymptotic normality of a single one of the $\widehat{\xi}_{\tau_n,j}$ has been investigated in Corollary~2 of~\cite{daogirstu2018}. To write the corresponding joint convergence result, we require the following set of second-order conditions designed to control the rate of convergence in~\eqref{eqn:expectiles_connection}.
\begin{cond}\label{cond:multivariate_bis}
Assume that Condition~\ref{cond:multivariate}\ref{cond:enu:reg_varying} holds and that, for every $1\leq j\leq d$, 
\[
\forall x>0, \ \lim_{s\to\infty}\frac{1}{A_j(s)}\left(\frac{U_j(sx)}{U_j(s)}-x^{\gamma_{j}}\right)=x^{\gamma_j}\frac{x^{\rho_j}-1}{\rho_j},
\]
where $\rho_j\leq 0$ and $A_j$ is a measurable function converging to 0 at infinity and having constant sign. Hereafter, $(x^{\rho_j}-1)/\rho_j$ is to be read as $\log(x)$ when $\rho_j=0$.
\end{cond}
Condition~\ref{cond:multivariate_bis} controls rates of convergences in Condition~\ref{cond:multivariate}\ref{cond:enu:reg_varying}: since $|A_j|$ is regularly varying with index $\rho_j$~\citep[by Theorems~2.3.3 and~2.3.9 in][]{haafer2006}, the larger $|\rho_j|$ is, the faster $|A_j|$ converges to~0 and the smaller the error in the approximation of the right tail of $U_j$ by a purely Pareto tail will be. 
Any distribution part of the Hall-Welsh class~\eqref{eqn:halwel} satisfies this kind of condition~\citep[as a consequence of Theorem 2.3.9 in][]{haafer2006}. Numerous examples of commonly used distributions that satisfy this assumption can be found in~\cite{beigoesegteu2004}. 
\vskip1ex
\noindent
Our next result, of interest in its own right, examines the joint convergence between Hill estimators and intermediate order statistics across marginals. 
A related result, 
limited to joint convergence of Hill estimators only, is Theorem~4 in~\cite{stu2019}. 
\begin{Theo}\label{theo:asy2ndorderinter}
Assume that Conditions~\ref{cond:multivariate} and~\ref{cond:multivariate_bis} hold. Let $\tau_n\uparrow 1$ be such that $n(1-\tau_n)\to\infty$ and, for any $1\leq j\leq d$, $\sqrt{n(1-\tau_n)}A_j((1-\tau_n)^{-1})\to\lambda_j\in\RR$ as $n\to\infty$. Then we have 
\[
\sqrt{n(1-\tau_n)} \left(\widehat{\gamma}_{\tau_n,j}-\gamma_j, \frac{\widehat{q}_{\tau_n,j}}{q_{\tau_n,j}} - 1 \right)_{1\leq j\leq d} \tod \mathcal{N}_{2d}\left((\lambda_j/(1-\rho_j),0)_{1\leq j\leq d}, \bfSigma^{\mathrm{Q}}(\bfgamma,\bfR) \right). 
\]
The covariance matrix $\bfSigma^{\mathrm{Q}}(\bfgamma,\bfR)$ can be partitioned into $2\times 2$ blocks $\bfSigma_{j,\ell}^{\mathrm{Q}}(\bfgamma,\bfR)$, given by $\bfSigma_{j,j}^{\mathrm{Q}}(\bfgamma,\bfR)=\gamma_j^2 \mathrm{I}_2$ (where $\mathrm{I}_2$ denotes the $2\times 2$ identity matrix) for any $j\in \{1,\ldots,d\}$ and 
\[
\bfSigma_{j,\ell}^{\mathrm{Q}}(\bfgamma,\bfR) = \gamma_j\gamma_{\ell} \left( \begin{array}{cc} R_{j,\ell}(1,1) & \displaystyle \int_0^1 R_{j,\ell}(u,1) \frac{\diff u}{u}-R_{j,\ell}(1,1) \\ \displaystyle \int_0^1 R_{j,\ell}(1,u) \frac{\diff u}{u}-R_{j,\ell}(1,1) & R_{j,\ell}(1,1) \end{array} \right)
\]
for any $j,\ell\in \{1,\ldots,d\}$ with $j<\ell$.
\end{Theo}
%
Let us highlight that 
although there is asymptotically no correlation between the Hill estimator for a given marginal and the corresponding order statistic~\citep[see also Lemma~3.2.3 p.71 in][]{haafer2006}, there are generally nonzero correlations between pairs of Hill estimators, pairs of intermediate order statistics, as well as between the Hill estimator of a given marginal and an intermediate order statistic pertaining to another marginal. 
%

The desired result on the joint convergence of the $\widehat{\xi}_{\tau_n,j}$ is now a corollary of Theorem \ref{theo:asy2ndorderinter}. Set $m(x)=(1-x)^{-1}-\log(x^{-1}-1)$, for $x\in (0,1)$.
\begin{Coro}\label{coro:asy2ndorder}
Work under the conditions of Theorem \ref{theo:asy2ndorderinter}. Assume in addition that $\EE|\min(X_j,0)|<\infty$, that $0<\gamma_j<1$ for any $1\leq j\leq d$ and that $\sqrt{n(1-\tau_n)}q^{-1}_{\tau_n,j}\to \mu_j\in\RR$ as $n\to\infty$. Then
\[
\sqrt{n(1-\tau_n)} \left(\frac{\widehat{\xi}_{\tau_n,j}}{\xi_{\tau_n,j}} - 1 \right)_{1\leq j\leq d} \tod \mathcal{N}_{d}\left(\bfb^{\mathrm{QB}}, \bfV^{\mathrm{QB}}(\bfgamma,\bfR) \right),
\]
where the asymptotic bias $\bfb^{\mathrm{QB}}$ has components
\[
\bfb_j^{\mathrm{QB}} = -\gamma_j (\gamma_j^{-1}-1)^{\gamma_j} \EE(X_j)\mu_j + \left( \frac{m(\gamma_j)}{1-\rho_j} - \frac{(\gamma_j^{-1}-1)^{-\rho_j}}{1-\gamma_j-\rho_j} - \dfrac{(\gamma_j^{-1}-1)^{-\rho_j}-1}{\rho_j} \right) \lambda_j
\]
and the covariance matrix $\bfV^{\mathrm{QB}}(\bfgamma,\bfR)$ has entries
\[
\bfV_{j,\ell}^{\mathrm{QB}}(\bfgamma,\bfR) =
\begin{cases} \gamma_j^2 (1+(m(\gamma_j))^2)& \mbox{if } j=\ell, \\[10pt]
\gamma_{j}\gamma_{\ell} \bigg( R_{j,\ell}(1,1)(m(\gamma_j)-1)(m(\gamma_\ell)-1) &  \\[10pt] 
 \qquad \ +\displaystyle m(\gamma_j)\int_0^1 R_{j,\ell}(u,1)\frac{\diff u}{u} + m(\gamma_\ell)\int_0^1 R_{j,\ell}(1,u)\frac{\diff u}{u}\bigg) & \mbox{if } j<\ell.
\end{cases}
\]
\end{Coro}
This result is the multivariate extension of Corollary~2 in~\cite{daogirstu2018} that is required for our purposes. Note also that unlike the latter, our result is written without the unnecessary assumption of an increasing (marginal) distribution function.

%
\subsection{At the extreme level}
\label{sec:multiple_extreme}
%
We consider now the problem of most relevance to risk management in practice, which is to estimate extreme expectiles $\xi_{\tau'_n,j}$, where $\tau'_n\to 1$ is such that $n(1-\tau'_n)\to c\in [0,\infty)$. In risk management, one would typically consider $\tau'_n\geq 1-1/n$, see for example Chapter~4 of~\cite{haafer2006} and~\cite{caieinhaazho2015} in the context of extreme quantile esstimation.  
The basic idea, dating back to~\cite{wei1978}, is to extrapolate intermediate expectile estimators at level $\tau_n$ to the extreme level $\tau'_n$, beyond the observed data, using the marginal heavy tails assumption. This is warranted by convergence~\eqref{eqn:expectiles_connection}, which entails 
\[
\frac{\xi_{\tau'_n,j}}{\xi_{\tau_n,j}} \approx \frac{q_{\tau'_n,j}}{q_{\tau_n,j}} = \frac{U_j((1-\tau'_n)^{-1})}{U_j((1-\tau_n)^{-1})} \approx \left( \frac{1-\tau'_n}{1-\tau_n} \right)^{-\gamma_j} \ \mbox{ as } n\to\infty. 
\]
This suggests 
the following two estimators: the LAWS-based extrapolating estimator
\[
\widetilde{\xi}_{\tau_n',j}^{\star} = \widetilde{\xi}_{\tau_n',j}^{\star}(\tau_n)=\left(\frac{1-\tau_n'}{1-\tau_n}\right)^{-\widehat{\gamma}_{\tau_n,j}} \widetilde{\xi}_{\tau_n,j}
\]
and the QB extrapolating estimator
\[
\widehat{\xi}_{\tau_n',j}^{\star} = \widehat{\xi}_{\tau_n',j}^{\star}(\tau_n)=\left(\frac{1-\tau_n'}{1-\tau_n}\right)^{-\widehat{\gamma}_{\tau_n,j}}\widehat{\xi}_{\tau_n,j}=(\widehat{\gamma}_{\tau_n,j}^{-1}-1)^{-\widehat{\gamma}_{\tau_n,j}}\widehat{q}^{\star}_{\tau'_n,j},
\]
where $\widehat{q}^{\star}_{\tau'_n,j}$ is the Weissman estimator of the extreme quantile $q_{\tau'_n,j}$~\citep{wei1978}. Our next main result towards our goal of carrying out joint inference about extreme expectiles is a statement of the joint convergence of these estimators across marginals.
\begin{Theo}
\label{theo:asyext}
Assume that Conditions~\ref{cond:multivariate} and \ref{cond:multivariate_bis} hold, with $\rho_j<0$ for any $j\in\{1,\ldots,d \}$. Let $\tau_n,\tau'_n\uparrow 1$ with $n(1-\tau_n)\to\infty$, $n(1-\tau'_n)\to c\in[0,\infty)$ and $\sqrt{n(1-\tau_n)}/\log[(1-\tau_n)/(1-\tau_n')]\to\infty$ as $n\to\infty$. Assume also that for any $1\leq j\leq d$, $\sqrt{n(1-\tau_n)}q^{-1}_{\tau_n,j}\to \mu_j\in\RR$ and $\sqrt{n(1-\tau_n)}A_j((1-\tau_n)^{-1})\to\lambda_j\in\RR$ as $n\to\infty$. Let $\bfb^{\star}=(\lambda_j/(1-\rho_j))_{1\leq j\leq d}$ and define a covariance matrix $\bfV^{\star}(\bfgamma,\bfR)$ by
\[
\bfV_{j,\ell}^{\star}(\bfgamma,\bfR) = \begin{cases} \gamma_j^2 & \mbox{if } j=\ell, \\
\gamma_{j}\gamma_{\ell} \, R_{j,\ell}(1,1) & \mbox{if } j<\ell.
\end{cases}
\]
\begin{enumerate}[label=(\roman*)]
\item Assume that there is $\delta>0$ such that $\EE|\min(X_j,0)|^{2+\delta}<\infty$ and that $0<\gamma_j<1/2$ for any $1\leq j\leq d$. Then 
\[
\frac{\sqrt{n(1-\tau_n)}}{\log[(1-\tau_n)/(1-\tau_n')]}\left(\frac{\widetilde{\xi}_{\tau_n',j}^{\star}}{\xi_{\tau'_n,j}}-1\right)_{1\leq j\leq d} \tod \mathcal{N}_{d}\left(\bfb^{\star}, \bfV^{\star}(\bfgamma,\bfR) \right).
\]
\item Assume that $\EE|\min(X_j,0)|<\infty$ and that $0<\gamma_j<1$ for any $1\leq j\leq d$. Then 
\[
\frac{\sqrt{n(1-\tau_n)}}{\log[(1-\tau_n)/(1-\tau_n')]}\left(\frac{\widehat{\xi}_{\tau_n',j}^{\star}}{\xi_{\tau'_n,j}}-1\right)_{1\leq j\leq d} \tod \mathcal{N}_{d}\left(\bfb^{\star}, \bfV^{\star}(\bfgamma,\bfR) \right).
\]
\end{enumerate}
\end{Theo}
This result generalises the convergence of a single one of either the $\widetilde{\xi}_{\tau_n',j}^{\star}$ or $\widehat{\xi}_{\tau_n',j}^{\star}$, examined in Corollaries~3 and~4 of~\cite{daogirstu2018}. It is proven by showing that the joint asymptotic Gaussian distribution of our Weissman-type extrapolating estimators is exclusively governed by that of the Hill estimators used in the extrapolation procedure. However, and even though the asymptotic behaviour of the Hill estimators is certainly crucial, correctly inferring the anchor intermediate expectile will also be important in finite-sample situations, as we shall show in our construction of confidence regions and in our simulation study. 

\section{Joint inference on extreme expectiles}\label{sec:CI} 

Equipped with our theory developed in Section~\ref{sec:main}, we derive asymptotic confidence regions for inference about extreme expectiles and provide a testing procedure for their equality. We start by the construction of confidence regions at intermediate and extreme levels. Of course, the study of the intermediate case is less important in practice since most applications in tail risk management focus on the estimation of risk measures at properly extreme levels. However, as we shall illustrate below, 
giving an accurate measure of the uncertainty about intermediate expectile estimators 
will be key to our definition of accurate Gaussian confidence regions for multiple extreme expectiles. 
Throughout this section, we let $\boldsymbol{\xi}_{\tau_n} = (\xi_{\tau_n,1},\ldots,\xi_{\tau_n,d})^{\top}$ and define similarly $\boldsymbol{\xi}_{\tau'_n}$, $\widetilde{\boldsymbol{\xi}}_{\tau_n}$, $\widehat{\boldsymbol{\xi}}_{\tau_n}$, $\bfxitilde_{\tau'_n}
^{\star}$ and $\bfxihat_{\tau'_n}^{\star}$. The symbol $\boldsymbol{1}_d$ denotes the $d-$dimensional vector with all entries equal to 1. All operations on vectors, apart from matrix operations, are meant componentwise. 
\subsection{Asymptotic confidence region construction: intermediate case}
\label{sec:CI_inter}

\noindent
{\bf Using LAWS estimation \hskip1ex} Our main instrument is Theorem~\ref{theo:asydirinter}, namely
\begin{equation}
\label{eqn:expctile_est_LAWS}
\sqrt{n(1-\tau_n)}\left( \frac{\widetilde{\boldsymbol{\xi}}_{\tau_n}}{\boldsymbol{\xi}_{\tau_n}} - \boldsymbol{1}_d \right) \tod \mathcal{N}_d( \bfzero_d, \bfV^{\mathrm{LAWS}}(\bfgamma,\bfR) ).
\end{equation}
%
Using this Gaussian asymptotic approximation to build a confidence region for $\boldsymbol{\xi}_{\tau_n}$ is a delicate task. In the multivariate case, this problem is even more difficult because of the additional nontrivial question of estimating the off-diagonal elements of $\bfV^{\mathrm{LAWS}}(\bfgamma,\bfR)$ to model correctly the dependence between LAWS estimators.

We investigate here a solution based on the proof of Theorem~\ref{theo:asydirinter}. If $\varphi_{\tau}(y) = | \tau-\ind\{ y\leq 0\} | y$ is the derivative of $\eta_{\tau}/2$, one has the following nonparametric approximation of $\bfV_{j,\ell}^{\mathrm{LAWS}}(\bfgamma,\bfR)$ for large $n$: 
\begin{align*}
 & \bfV_{j,\ell}^{\mathrm{LAWS}}(\bfgamma,\bfR) \\
 &\approx \frac{1}{(1-\tau_n) \xi_{\tau_n,j} \xi_{\tau_n,\ell}} \times \frac{\EE( \varphi_{\tau_n}(X_j-\xi_{\tau_n,j}) \varphi_{\tau_n}(X_\ell-\xi_{\tau_n,\ell}) )}{[1+(2\tau_n-1)\overline{F}_j(\xi_{\tau_n,j})/(1-\tau_n)][1+(2\tau_n-1)\overline{F}_{\ell}(\xi_{\tau_n,\ell})/(1-\tau_n)]}.
\end{align*}
This approximation is our starting point for the construction of an estimator of $\bfV_{j,\ell}^{\mathrm{LAWS}}(\bfgamma,\bfR)$. One could estimate each term in this nonparametric approximation directly; this turns out not to be the best-performing solution in practice because it tends to provide an underestimation of the marginal uncertainty on expectiles. Our solution, suggested by the results of extensive Monte-Carlo simulations, is the following. For the diagonal entry $\bfV_{j,j}^{\mathrm{LAWS}}(\bfgamma,\bfR)$ of $\bfV^{\mathrm{LAWS}}(\bfgamma,\bfR)$, one has
\[
\bfV_{j,j}^{\mathrm{LAWS}}(\bfgamma,\bfR)\approx \frac{2\gamma_j^2}{1-2\gamma_j} \times \frac{1+\overline{F}_j(\xi_{\tau_n,j})/(1-\tau_n)}{[1+(2\tau_n-1)\overline{F}_j(\xi_{\tau_n,j})/(1-\tau_n)]^2} \ \mbox{for large } n. 
\]
For off-diagonal elements, the covariance $\operatorname{Cov}( \varphi_{\tau_n}(X_j-\xi_{\tau_n,j}), \varphi_{\tau_n}(X_\ell-\xi_{\tau_n,\ell}) )$ is in practice found to be a good approximation of the direction of dependence within the data; a finite-sample improvement on the estimation of the strength of this dependence is found by writing 
\[
\bfV_{j,\ell}^{\mathrm{LAWS}}(\bfgamma,\bfR) \approx \gamma_j\gamma_\ell \frac{\EE( \varphi_{\tau_n}(X_j-\xi_{\tau_n,j}) \varphi_{\tau_n}(X_\ell-\xi_{\tau_n,\ell}) )}{(1-\tau_n) \xi_{\tau_n,j} \xi_{\tau_n,\ell}} \ \mbox{for large } n. 
\]
Our estimator of $\bfV^{\mathrm{LAWS}}(\bfgamma,\bfR)$ is now constructed by plugging in the LAWS and Hill estimators, the empirical survival functions $\widehat{\overline{F}}_{n,j}$ based on the $X_{i,j}$ ($1\leq i\leq n$) 
and the empirical covariances
\[
\overline{m}_{n,j,\ell} = \frac{1}{n} \sum_{i=1}^n \varphi_{\tau_n}(X_{i,j}-\widetilde{\xi}_{\tau_n,j}) \varphi_{\tau_n}(X_{i,\ell}-\widetilde{\xi}_{\tau_n,\ell}).
\]
This results in the estimator $\bfVhat_n^{\mathrm{LAWS}}(\bfgamma,\bfR)$ of $\bfV^{\mathrm{LAWS}}(\bfgamma,\bfR)$ given elementwise by 
\begin{align*}
\bfVhat_{n,j,j}^{\mathrm{LAWS}}(\bfgamma,\bfR) &= \frac{2\widehat{\gamma}_{\tau_n,j}^2}{1-2\widehat{\gamma}_{\tau_n,j}} \times \frac{1+\widehat{\overline{F}}_{n,j}(\widetilde{\xi}_{\tau_n,j})/(1-\tau_n)}{\left[ 1+(2\tau_n-1) \widehat{\overline{F}}_{n,j}(\widetilde{\xi}_{\tau_n,j})/(1-\tau_n) \right]^2} \\
\mbox{and } \bfVhat_{n,j,\ell}^{\mathrm{LAWS}}(\bfgamma,\bfR) &= \widehat{\gamma}_{\tau_n,j} \widehat{\gamma}_{\tau_n,\ell} \frac{\overline{m}_{n,j,\ell}}{(1-\tau_n)\widetilde{\xi}_{\tau_n,j} \widetilde{\xi}_{\tau_n,\ell}} \mbox{ for } j\neq \ell.
\end{align*}
Under the assumptions of Theorem~\ref{theo:asydirinter}, this is indeed a consistent estimator of $\bfV^{\mathrm{LAWS}}(\bfgamma,\bfR)$. When $\bfV^{\mathrm{LAWS}}(\bfgamma,\bfR)$ is symmetric positive definite (in particular, no perfect asymptotic dependence between two components of $\bfX$ can be present), 
multiplying the left-hand side in~\eqref{eqn:expctile_est_LAWS} by the positive definite inverse square root $[ \bfV^{\mathrm{LAWS}}(\bfgamma,\bfR) ]^{-1/2}$ of $\bfV^{\mathrm{LAWS}}(\bfgamma,\bfR)$ and then plugging in our estimator $\bfVhat_n^{\mathrm{LAWS}}(\bfgamma,\bfR)$ produces an asymptotically Gaussian random vector with independent standard Gaussian components. Therefore, if $\| \cdot \|_2$ denotes the Euclidean norm on $\RR^d$ and $\chi^2_{d,1-\alpha}$ denotes the $(1-\alpha)-$quantile of the chi-square distribution with $d$ degrees of freedom, one has 
\[
\mathbb{P}\left( \left\| \sqrt{n(1-\tau_n)} \left[ \bfVhat_n^{\mathrm{LAWS}}(\bfgamma,\bfR) \right]^{-1/2} \left( \frac{\widetilde{\boldsymbol{\xi}}_{\tau_n}}{\boldsymbol{\xi}_{\tau_n}} - \boldsymbol{1}_d \right) \right\|_2^2 \leq \chi^2_{d,1-\alpha} \right) \to 1-\alpha \ \mbox{ as } n\to\infty.
\]
Denoting by $B_d(\bfzero_d,r)$ the closed Euclidean ball in $\mathbb{R}^d$ whose centre is the origin and radius is $r$, we find the corresponding $(1-\alpha)-$asymptotic LAWS-based confidence region for $\boldsymbol{\xi}_{\tau_n}$ as the random ellipsoid
%
%
%
\begin{align*}
  \widetilde{\mathcal{E}}_{\tau_n,\alpha}  &= \widetilde{\boldsymbol{\xi}}_{\tau_n} \left[ \boldsymbol{1}_d +  \left[\bfVhat_n^{\mathrm{LAWS}}(\bfgamma,\bfR)\right]^{1/2} B_d\left(\bfzero_d,\sqrt{\chi^2_{d,1-\alpha}/n(1-\tau_n)} \right) \right] \\
	&= \left\{ \mathbf{z}\in \mathbb{R}^d \, | \, \exists \mathbf{u}\in B_d\left(\bfzero_d,\sqrt{\chi^2_{d,1-\alpha}/n(1-\tau_n)} \right), \ \mathbf{z} = \widetilde{\boldsymbol{\xi}}_{\tau_n} \left[ \boldsymbol{1}_d +  \left[\bfVhat_n^{\mathrm{LAWS}}(\bfgamma,\bfR)\right]^{1/2} \mathbf{u} \right] \right\}. 
\end{align*}
[Recall that all operations except the matrix product $\left[\bfVhat_n^{\mathrm{LAWS}}(\bfgamma,\bfR)\right]^{1/2} \mathbf{u}$ are meant componentwise.]
\vskip1ex
\noindent
{\bf Using QB estimation} With the QB estimator, our main tool is Corollary~\ref{coro:asy2ndorder}: 
\[
\sqrt{n(1-\tau_n)} \left(\frac{\widehat{\boldsymbol{\xi}}_{\tau_n}}{\boldsymbol{\xi}_{\tau_n}} - 1 \right) \tod \mathcal{N}_{d}\left(\bfb^{\mathrm{QB}}, \bfV^{\mathrm{QB}}(\bfgamma,\bfR) \right).
\]
Similarly to what is observed when using LAWS estimators, great care has to be taken in constructing confidence regions based on this convergence. 

Contrary to the LAWS estimator, the QB estimator is asymptotically biased due to its reliance on the relationship~\eqref{eqn:expectiles_connection}. The $j$th component of this bias is essentially, as $n\to\infty$, 
\begin{align*}
 \bfb_j^{\mathrm{QB}}   &\approx -\gamma_j (\gamma_j^{-1}-1)^{\gamma_j} \EE(X_j) \frac{\sqrt{n(1-\tau_n)}}{q_{\tau_n,j}} \\
                        &+ \left( \frac{m(\gamma_j)}{1-\rho_j} - \frac{(\gamma_j^{-1}-1)^{-\rho_j}}{1-\gamma_j-\rho_j} - \dfrac{(\gamma_j^{-1}-1)^{-\rho_j}-1}{\rho_j} \right) \sqrt{n(1-\tau_n)} A_j((1-\tau_n)^{-1}). 
\end{align*}
Two sources of bias therefore arise when using the QB estimator: one due to marginal tail heaviness and the other to the second-order framework. The correction of the latter source of bias involves estimating accurately the second-order parameter $\rho_j$, which is a notoriously difficult problem~\citep[see {\it e.g.}~the Introduction of][]{caihaazho2013}, especially from the practical point of view since consistent estimators of $\rho_j$ typically suffer from low rates of convergence, see {\it e.g.}~\citet[][p.2638]{goebeiwet2010} and~\citet[][p.298]{gompescae2009}. As such, correcting second-order bias tends to increase finite-sample variability substantially, resulting in confidence regions that may be too conservative. By contrast, the simple expression of the bias component proportional to $q_{\tau_n,j}$ makes its correction a straightforward task, with all estimators involved converging at the rate $\sqrt{n(1-\tau_n)}$ or more. This constitutes our rationale for concentrating specifically on the first source of bias with the estimator 
\[
\bfbhat_j^{\mathrm{QB}} = -\widehat{\gamma}_{\tau_n,j} (\widehat{\gamma}_{\tau_n,j}^{-1}-1)^{\widehat{\gamma}_{\tau_n,j}} \overline{X}_{n,j} \frac{\sqrt{n(1-\tau_n)}}{\widehat{q}_{\tau_n,j}}, \ \mbox{ where } \overline{X}_{n,j} = \frac{1}{n} \sum_{i=1}^n X_{i,j}. 
\]
The covariance matrix $\bfV^{\mathrm{QB}}(\bfgamma,\bfR)$, meanwhile, is estimated as follows: 
\begin{align*}
\bfVhat_{n,j,j}^{\mathrm{QB}}(\bfgamma,\bfR) &= \widehat{\gamma}_{\tau_n,j}^2 (1+(m(\widehat{\gamma}_{\tau_n,j}))^2) \ \mbox{ (with } m(x)=(1-x)^{-1}-\log(x^{-1}-1)) \\ 
\mbox{and } \bfVhat_{n,j,\ell}^{\mathrm{QB}}(\bfgamma,\bfR) &= \widehat{\gamma}_{\tau_n,j}\widehat{\gamma}_{\tau_n,\ell}\bigg( \widehat{R}_{\tau_n,j,\ell}(1,1)(m(\widehat{\gamma}_{\tau_n,j})-1)(m(\widehat{\gamma}_{\tau_n,\ell})-1) \\
    & \qquad \qquad \qquad + m(\widehat{\gamma}_{\tau_n,j})\int_0^1 \widehat{R}_{\tau_n,j,\ell}(u,1)\frac{\diff u}{u} + m(\widehat{\gamma}_{\tau_n,\ell})\int_0^1 \widehat{R}_{\tau_n,j,\ell}(1,u)\frac{\diff u}{u} \bigg)
\end{align*}
where the estimator of the tail copula function $R_{j,\ell}$ is defined as
\begin{equation}
\label{eqn:tail_copula_est}
\widehat{R}_{\tau_n,j,\ell}(u,v) = \frac{1}{n(1-\tau_n)}\sum_{i=1}^n\ind
\left\{
\frac{n+1 - r_{n,i,j}}{(n+1)(1-\tau_n)}\leq u, \frac{n+1 - r_{n,i,\ell}}{(n+1)(1-\tau_n)}\leq v
\right\}.
\end{equation}
%
%
[Here $r_{n,i,j}$ denotes the marginal rank of observation $X_{i,j}$.] This estimator is a slightly modified version of the estimator of the empirical upper tail copula estimator given in Equation~(13) in~\cite{schsta2006}. 
%
The estimator $\bfVhat_n^{\mathrm{QB}}(\bfgamma,\bfR)$ is a consistent estimator of $\bfV^{\mathrm{QB}}(\bfgamma,\bfR)$, by a combination of Theorem~\ref{theo:asy2ndorderinter} and known results on the uniform consistency of $\widehat{R}_{\tau_n,j,\ell}$, see~\citet[][Section~5]{schsta2006}. 
%
\vskip1ex
\noindent
A calculation entirely similar to the one carried out with the LAWS estimator now yields an $(1-\alpha)-$asymptotic QB confidence region for $\boldsymbol{\xi}_{\tau_n}$ as the random ellipsoid
\[
\widehat{\mathcal{E}}_{\tau_n,\alpha} = \widehat{\boldsymbol{\xi}}_{\tau_n} \left[ \boldsymbol{1}_d - \frac{\bfbhat^{\mathrm{QB}}}{\sqrt{n(1-\tau_n)}} + \left[\bfVhat_n^{\mathrm{QB}}(\bfgamma,\bfR)\right]^{1/2} B_d\left(\bfzero_d,\sqrt{\chi^2_{d,1-\alpha}/n(1-\tau_n)} \right) \right]. 
\]
A comparison of these regions in terms of actual coverage will be carried out in Section~\ref{sec:simulations}.

\subsection{Asymptotic confidence region construction: extreme case}
\label{sec:CI_ext}

At the extreme level, the key result for our purposes is Theorem~\ref{theo:asyext}. Nevertheless, if one constructs an asymptotic confidence region directly from this result, the actual finite-sample coverage probability can be quite poor, even in the estimation of a single extreme expectile: see Appendix A.2 in~\cite{daogirstu2018} where Gaussian QQ-plots show that the observed variance of extreme expectile estimators can be fairly different from the asymptotic variance in the Gaussian approximation. We shall illustrate this in more detail in Section~\ref{sec:simulations:1D}.

Our idea is to, first, get a finer understanding of the uncertainty in the estimation of extreme expectiles. The gist of our method is that any estimator of the form 
\[
\overline{\xi}_{\tau'_n,j}^{\star} = \left( \frac{1-\tau'_n}{1-\tau_n} \right)^{-\widehat{\gamma}_{\tau_n,j}} \overline{\xi}_{\tau_n,j}, 
\]
where $\overline{\xi}_{\tau_n,j}$ is a consistent estimator of $\xi_{\tau_n,j}$, satisfies
\begin{equation}
\label{eqn:weissrep}
\log\left( \frac{\overline{\xi}_{\tau'_n,j}^{\star}}{\xi_{\tau'_n,j}} \right) = (\widehat{\gamma}_{\tau_n,j}-\gamma_j) \log\left( \frac{1-\tau_n}{1-\tau'_n} \right) + \log\left( \frac{\overline{\xi}_{\tau_n,j}}{\xi_{\tau_n,j}} \right) - \log\left( \left[ \frac{1-\tau'_n}{1-\tau_n} \right]^{\gamma_j} \frac{\xi_{\tau'_n,j}}{\xi_{\tau_n,j}}  \right).
\end{equation}
Under the conditions of Theorem~\ref{theo:asyext}, the second (random) term and the third (bias) term are dominated by the first term, leading to the common asymptotic distribution obtained therein. 
In practice however, the behaviour of $\overline{\xi}_{\tau_n,j}$ matters, and so does the correlation between $\overline{\xi}_{\tau_n,j}$ and $\widehat{\gamma}_{\tau_n,j}$, especially when $\log d_n = \log[(1-\tau_n)/(1-\tau_n')]$ is only moderately large. Investigating this uncertainty and correlation will lead us to define corrected Gaussian asymptotic confidence regions. All our confidence regions will be constructed on the log-scale; using this scale has been shown to improve finite-sample coverage of confidence regions for extreme risk measures~\citep[see {\it e.g.}~p.628 in][in the context of extreme quantile estimation]{dre2003}. We found from Monte-Carlo simulations that this is also the case for expectiles. 
\vskip1ex
\noindent
{\bf Using the LAWS-based extrapolating estimator} The crucial result is an extension of Theorem~\ref{theo:asydirinter} giving the joint convergence of the Hill estimators and intermediate LAWS expectile estimators across marginals. 
\begin{Theo}\label{theo:asyinterjoint}
Assume that Conditions~\ref{cond:multivariate} and~\ref{cond:multivariate_bis} hold. Assume further that there is $\delta>0$ such that $\EE|\min(X_j,0)|^{2+\delta}<\infty$ and that $0<\gamma_j<1/2$ for any $1\leq j\leq d$. Let $\tau_n\uparrow 1$ be such that $n(1-\tau_n)\to\infty$ and, for any $1\leq j\leq d$, $\sqrt{n(1-\tau_n)}A_j((1-\tau_n)^{-1})\to\lambda_j\in\RR$ as $n\to\infty$. Then we have 
\[
\sqrt{n(1-\tau_n)} \left(\widehat{\gamma}_{\tau_n,j}-\gamma_j, \frac{\widetilde{\xi}_{\tau_n,j}}{\xi_{\tau_n,j}} - 1 \right)_{1\leq j\leq d} \tod \mathcal{N}_{2d}\left((\lambda_j/(1-\rho_j),0)_{1\leq j\leq d}, \bfSigma^{\mathrm{LAWS}}(\bfgamma,\bfR) \right). 
\]
The covariance matrix $\bfSigma^{\mathrm{LAWS}}(\bfgamma,\bfR)$ is partitioned into $2\times 2$ blocks $\bfSigma^{\mathrm{LAWS}}_{j,\ell}(\bfgamma,\bfR)$ given by 
\[
\bfSigma^{\mathrm{LAWS}}_{j,j}(\bfgamma,\bfR)=\gamma_j^2\left( \begin{array}{cc} 1 & \dfrac{\gamma_j(\gamma_j^{-1}-1)^{\gamma_j}}{(1-\gamma_j)^2} \\[15pt] \dfrac{\gamma_j(\gamma_j^{-1}-1)^{\gamma_j}}{(1-\gamma_j)^2} & \dfrac{2\gamma_j}{1-2\gamma_j} \end{array} \right)
\]
when $j=\ell\in \{1,\ldots,d\}$ and, elementwise,
\begin{align*}
\bfSigma^{\mathrm{LAWS}}_{j,\ell}(\bfgamma,\bfR)(1,1) &= \gamma_j\gamma_{\ell} R_{j,\ell}(1,1), \\
\bfSigma^{\mathrm{LAWS}}_{j,\ell}(\bfgamma,\bfR)(1,2) &= \gamma_{\ell}\iint_{[1,\infty)^2} R_{j,\ell}\left(x_j^{-1/\gamma_j},(\gamma_\ell^{-1}-1) x_{\ell}^{-1/\gamma_\ell} \right) \frac{\diff x_j}{x_j} \, \diff x_{\ell} \\
	 &- \gamma_j\gamma_{\ell}\int_1^{\infty} R_{j,\ell}\left(1,(\gamma_\ell^{-1}-1) x_{\ell}^{-1/\gamma_\ell} \right) \diff x_{\ell}, \\
\bfSigma^{\mathrm{LAWS}}_{j,\ell}(\bfgamma,\bfR)(2,1) &= \gamma_j\iint_{[1,\infty)^2} R_{j,\ell}\left((\gamma_j^{-1}-1) x_j^{-1/\gamma_j},x_{\ell}^{-1/\gamma_{\ell}} \right) \diff x_j \, \frac{\diff x_{\ell}}{x_{\ell}} \\
	 &- \gamma_j\gamma_{\ell}\int_1^{\infty} R_{j,\ell}\left((\gamma_j^{-1}-1) x_j^{-1/\gamma_j},1 \right) \diff x_j, \\
\bfSigma^{\mathrm{LAWS}}_{j,\ell}(\bfgamma,\bfR)(2,2) &= \gamma_j\gamma_{\ell} \iint_{[1,\infty)^2} R_{j,\ell}\left((\gamma_j^{-1}-1) x_j^{-1/\gamma_j},(\gamma_\ell^{-1}-1) x_{\ell}^{-1/\gamma_\ell} \right) \diff x_j \, \diff x_{\ell}
\end{align*}
for any $j,\ell\in \{1,\ldots,d\}$ with $j<\ell$.
\end{Theo}
Theorem~\ref{theo:asyinterjoint} and Equation~\eqref{eqn:weissrep} suggest the following approximation for the LAWS-based extrapolating estimator on the log-scale: 
\[
\frac{\sqrt{n(1-\tau_n)}}{\log[(1-\tau_n)/(1-\tau_n')]} \left[ \log \left( \frac{\bfxitilde_{\tau_n'}^{\star}}{\boldsymbol{\xi}_{\tau'_n}} \right) + \log\left( \left[ \frac{1-\tau'_n}{1-\tau_n} \right]^{\boldsymbol{\gamma}} \frac{\boldsymbol{\xi}_{\tau'_n}}{\boldsymbol{\xi}_{\tau_n}}  \right) \right] \approx \mathcal{N}_{d}\left(\bfzero_d, \bfV_n^{\star,\mathrm{LAWS}}(\bfgamma,\bfR) \right),
\]
where $\bfV_n^{\star,\mathrm{LAWS}}(\bfgamma,\bfR)$ is defined elementwise as (recall that $\log d_n = \log[(1-\tau_n)/(1-\tau_n')]$) 
\[
\bfV_{n,j,\ell}^{\star,\mathrm{LAWS}}(\bfgamma,\bfR) = \left( \begin{array}{c} 1 \\ 1/\log d_n \end{array} \right)^{\top} \bfSigma^{\mathrm{LAWS}}_{j,\ell}(\bfgamma,\bfR) \left( \begin{array}{c} 1 \\ 1/\log d_n \end{array} \right).
\]
We now focus on the estimation of the bias term appearing in the above distributional approximation, and of the matrix $\bfV_n^{\star,\mathrm{LAWS}}(\bfgamma,\bfR)$. Use Proposition~1(i) in~\cite{daogirstu2020} and the proof of Theorem~4.3.8 in~\cite{haafer2006} to find 
\begin{align*}
-\log\left( \left[ \frac{1-\tau'_n}{1-\tau_n} \right]^{\gamma_j} \frac{\xi_{\tau'_n,j}}{\xi_{\tau_n,j}}  \right) 
 &\approx \frac{\gamma_j (\gamma_j^{-1}-1)^{\gamma_j} \EE(X_j)}{q_{\tau_n,j}} + \operatorname{O}(A_j((1-\tau_n)^{-1})). 
\end{align*}
Here and as above we neither emphasise nor estimate the bias term proportional to $A_j((1-\tau_n)^{-1})$. 
We therefore suggest the following working approximation: 
\[
-\log\left( \left[ \frac{1-\tau'_n}{1-\tau_n} \right]^{\gamma_j} \frac{\xi_{\tau'_n,j}}{\xi_{\tau_n,j}}  \right) \approx \frac{\gamma_j (\gamma_j^{-1}-1)^{\gamma_j} \EE(X_j)}{q_{\tau_n,j}}
\]
which can be estimated by $-\bfbhat_j^{\mathrm{QB}}/\sqrt{n(1-\tau_n)}$ (see Section~\ref{sec:CI_inter}). To find an estimator of the covariance matrix $\bfV_n^{\star,\mathrm{LAWS}}(\bfgamma,\bfR)$, we note that
\begin{align*}
 \bfSigma^{\mathrm{LAWS}}_{j,\ell}(\bfgamma,\bfR)(1,1) &= \begin{cases} \gamma_j^2 & \mbox{if } j=\ell \\ \gamma_j\gamma_{\ell} R_{j,\ell}(1,1) & \mbox{if } j<\ell \end{cases} \\ 
 \mbox{is estimated by } \bfSigmahat^{\mathrm{LAWS}}_{n,j,\ell}(\bfgamma,\bfR)(1,1) &= \begin{cases} \widehat{\gamma}_{\tau_n,j}^2 & \mbox{if } j=\ell \\ \widehat{\gamma}_{\tau_n,j}\widehat{\gamma}_{\tau_n,\ell} \widehat{R}_{\tau_n,j,\ell}(1,1) & \mbox{if } j<\ell \end{cases}
\end{align*}
with the notation of Section~\ref{sec:CI_inter}. Similarly $\bfSigma^{\mathrm{LAWS}}_{j,\ell}(\bfgamma,\bfR)(2,2) = \bfV^{\mathrm{LAWS}}_{j,\ell}(\bfgamma,\bfR)$ is estimated with $\bfVhat^{\mathrm{LAWS}}_{n,j,\ell}(\bfgamma,\bfR)$. 
An estimation method for the off-diagonal entry $\bfSigma^{\mathrm{LAWS}}_{j,\ell}(\bfgamma,\bfR)(1,2)$ is obtained by noting that
\begin{align*}
\bfSigma^{\mathrm{LAWS}}_{j,\ell}(\bfgamma,\bfR)(1,2)	&\approx \gamma_j \gamma_{\ell} \frac{\operatorname{Cov}\left( [\log X_j - \log q_{\tau_n,j}] \ind{\{ X_j>q_{\tau_n,j} \}}, \varphi_{\tau_n}(X_\ell-\xi_{\tau_n,\ell}) \right)}{\mathbb{E}([\log X_j - \log q_{\tau_n,j}] \ind{\{ X_j>q_{\tau_n,j} \}}) \xi_{\tau_n,\ell}}  \\
	&-\gamma_j \gamma_{\ell} \frac{\operatorname{Cov}\left( \ind{\{ X_j>q_{\tau_n,j} \}}, \varphi_{\tau_n}(X_\ell-\xi_{\tau_n,\ell}) \right)}{\mathbb{P}(X_j>q_{\tau_n,j}) \xi_{\tau_n,\ell}}.
\end{align*}
We thus estimate $\bfSigma^{\mathrm{LAWS}}_{j,\ell}(\bfgamma,\bfR)(1,2)$ with $\bfSigmahat^{\mathrm{LAWS}}_{n,j,j}(\bfgamma,\bfR)(1,2) = \widehat{\gamma}_{\tau_n,j}^3 (\widehat{\gamma}_{\tau_n,j}^{-1}-1)^{\widehat{\gamma}_{\tau_n,j}}/(1-\widehat{\gamma}_{\tau_n,j})^2$ when $j=\ell$, and otherwise by (recall that $\varphi_{\tau_n}(X_\ell-\xi_{\tau_n,\ell})$ has expectation 0):
\begin{align*}
 & \bfSigmahat^{\mathrm{LAWS}}_{n,j,\ell}(\bfgamma,\bfR)(1,2) \\
 &= \frac{\widehat{\gamma}_{\tau_n,\ell}}{(1-\tau_n) \widetilde{\xi}_{\tau_n,\ell}} \, \frac{1}{n} \sum_{i=1}^n [\log X_{i,j} - \log X_{n-\lfloor n(1-\tau_n) \rfloor,n,j}] \ind{\{ X_{i,j}>X_{n-\lfloor n(1-\tau_n) \rfloor,n,j} \}} \varphi_{\tau_n}(X_{i,\ell}-\widetilde{\xi}_{\tau_n,\ell}) \\
 &- \frac{\widehat{\gamma}_{\tau_n,j}\widehat{\gamma}_{\tau_n,\ell}}{(1-\tau_n) \widetilde{\xi}_{\tau_n,\ell}} \, \frac{1}{n} \sum_{i=1}^n \ind{\{ X_{i,j}>X_{n-\lfloor n(1-\tau_n) \rfloor,n,j} \}} \varphi_{\tau_n}(X_{i,\ell}-\widetilde{\xi}_{\tau_n,\ell}).
\end{align*}
The entry $\bfSigma^{\mathrm{LAWS}}_{j,\ell}(\bfgamma,\bfR)(2,1)$ is estimated by $\bfSigmahat^{\mathrm{LAWS}}_{n,j,\ell}(\bfgamma,\bfR)(2,1)$ defined in a similar fashion by exchanging $j$ and $\ell$. This suggests an estimator of $\bfV_n^{\star,\mathrm{LAWS}}(\bfgamma,\bfR)$ defined elementwise as 
\[
\bfVhat_{n,j,\ell}^{\star,\mathrm{LAWS}}(\bfgamma,\bfR) = \left( \begin{array}{c} 1 \\ 1/\log d_n \end{array} \right)^{\top} \bfSigmahat^{\mathrm{LAWS}}_{n,j,\ell}(\bfgamma,\bfR) \left( \begin{array}{c} 1 \\ 1/\log d_n \end{array} \right).
\]
We finally deduce an $(1-\alpha)-$asymptotic LAWS-based confidence region for the extreme expectile $\boldsymbol{\xi}_{\tau'_n}$ as the deformed random ellipsoid
\[
\widetilde{\mathcal{E}}_{\tau'_n,\alpha}^{\star} = \bfxitilde_{\tau'_n}^{\star} \exp\Bigg( \frac{\bfbhat^{\mathrm{QB}}}{\sqrt{n(1-\tau_n)}} + \left[ \bfVhat_n^{\star,\mathrm{LAWS}}(\bfgamma,\bfR) \right]^{1/2} B_d\left(\bfzero_d,\sqrt{\chi^2_{d,1-\alpha}/n(1-\tau_n)} \, \log d_n \right) \Bigg).
\]
One can easily deduce from that construction a LAWS-based asymptotic $(1-\alpha)-$confidence interval for the $j$th marginal extreme expectile $\xi_{\tau'_n,j}$:
\begin{align*}
\widetilde{I}_{\tau'_n,j,\alpha}^{\star}    &= \left[ \widetilde{\xi}_{\tau'_n,j}^{\star} \exp\left( \frac{\bfbhat_j^{\mathrm{QB}}}{\sqrt{n(1-\tau_n)}} - \frac{\log d_n}{\sqrt{n(1-\tau_n)}} \sqrt{\bfVhat_{n,j,j}^{\star,\mathrm{LAWS}}(\bfgamma,\bfR)} z_{1-\alpha/2}  \right), \right. \\
                                        & \qquad \left. \widetilde{\xi}_{\tau'_n,j}^{\star} \exp\left( \frac{\bfbhat_j^{\mathrm{QB}}}{\sqrt{n(1-\tau_n)}} + \frac{\log d_n}{\sqrt{n(1-\tau_n)}} \sqrt{\bfVhat_{n,j,j}^{\star,\mathrm{LAWS}}(\bfgamma,\bfR)} z_{1-\alpha/2}  \right) \right]
\end{align*}
where $z_{1-\alpha/2}$ is the quantile of the standard Gaussian distribution at level $1-\alpha/2$. This can be seen as an adjusted version of the confidence interval based on the LAWS estimator that is considered in~\cite{daogirstu2018}. 
\vskip1ex
\noindent
{\bf Using the QB extrapolating estimator} We rewrite Equation~\eqref{eqn:weissrep} for $\widehat{\xi}_{\tau'_n,j}^{\star}$ as
\begin{align*}
\log\left( \frac{\widehat{\xi}_{\tau'_n,j}^{\star}}{\xi_{\tau'_n,j}} \right) &= (\widehat{\gamma}_{\tau_n,j}-\gamma_j) \log\left( \frac{1-\tau_n}{1-\tau'_n} \right) + \log\left( \frac{(\widehat{\gamma}^{-1}_{\tau_n,j}-1)^{-\widehat{\gamma}_{\tau_n,j}}}{(\gamma^{-1}_j-1)^{-\gamma_j}} \right) + \log\left( \frac{\widehat{q}_{\tau_n,j}}{q_{\tau_n,j}} \right) \\
 &- \log\left( \frac{\xi_{\tau'_n,j}}{(\gamma^{-1}_j-1)^{-\gamma_j} q_{\tau'_n,j}} \right) - \log\left( \left[ \frac{1-\tau'_n}{1-\tau_n} \right]^{\gamma_j} \frac{q_{\tau'_n,j}}{q_{\tau_n,j}}  \right). 
\end{align*}
%
By Proposition~1(i) in~\cite{daogirstu2020}, the first component of the bias on the second line of the right-hand side is essentially a linear combination of $1/q_{\tau'_n,j}$ and $A_j((1-\tau'_n)^{-1})$, which at the extreme level $\tau'_n$ are typically very small. The second component, meanwhile, is asymptotically proportional to $A_j((1-\tau_n)^{-1})$~\cite[see the proof of Theorem~4.3.8 of][]{haafer2006}, and we have discussed previously how estimating this kind of bias component is not necessarily beneficial for confidence region construction. We then ignore these two bias terms and use a Taylor expansion to write, as $n\to\infty$, 
\[
\log\left( \frac{\widehat{\xi}_{\tau'_n,j}^{\star}}{\xi_{\tau'_n,j}} \right) \approx (\widehat{\gamma}_{\tau_n,j}-\gamma_j) \left( m(\gamma_j) + \log d_n \right) + \log\left( \frac{\widehat{q}_{\tau_n,j}}{q_{\tau_n,j}} \right) + \operatorname{o}_{\PP}\left( \frac{1}{\sqrt{n(1-\tau_n)}} \right). 
\]
Using Theorem~\ref{theo:asy2ndorderinter} suggests the following approximation for the QB extrapolating estimator: 
\[
\frac{\sqrt{n(1-\tau_n)}}{\log[(1-\tau_n)/(1-\tau_n')]} \log \left( \frac{\bfxihat_{\tau_n'}^{\star}}{\boldsymbol{\xi}_{\tau'_n}} \right) \approx \mathcal{N}_{d}\left(\bfzero_d, \bfV_n^{\star,\mathrm{QB}}(\bfgamma,\bfR) \right),
\]
as $n\to\infty$, where $\bfV_n^{\star,\mathrm{QB}}(\bfgamma,\bfR)$ is defined elementwise as 
\[
\bfV_{n,j,\ell}^{\star,\mathrm{QB}}(\bfgamma,\bfR) =
\begin{cases} \dfrac{\gamma_j^2}{(\log d_n)^2} (1+(m(\gamma_j)+\log d_n)^2) & \mbox{if } j=\ell, \\[10pt]
\dfrac{\gamma_{j}\gamma_{\ell}}{(\log d_n)^2}\bigg(R_{j,\ell}(1,1)(m(\gamma_j)+\log (d_n)-1)(m(\gamma_\ell)+\log (d_n)-1) &  \\[10pt] 
 +\displaystyle (m(\gamma_j)+\log d_n)\int_0^1 R_{j,\ell}(u,1)\frac{\diff u}{u} + (m(\gamma_\ell)+\log d_n)\int_0^1 R_{j,\ell}(1,u)\frac{\diff u}{u}\bigg) & \mbox{if } j<\ell.
\end{cases}
\]
This matrix is readily estimated with the matrix $\bfVhat_n^{\star,\mathrm{QB}}(\bfgamma,\bfR)$ defined as 
\begin{align*}
 & \bfVhat_{n,j,\ell}^{\star,\mathrm{QB}}(\bfgamma,\bfR) \\
 &= \begin{cases} \dfrac{\widehat{\gamma}_{\tau_n,j}^2}{(\log d_n)^2} (1+[m(\widehat{\gamma}_{\tau_n,j})+\log d_n]^2) & \mbox{if } j=\ell, \\[10pt]
\dfrac{\widehat{\gamma}_{\tau_n,j}\widehat{\gamma}_{\tau_n,\ell}}{(\log d_n)^2}\bigg(\widehat{R}_{\tau_n,j,\ell}(1,1)(m(\widehat{\gamma}_{\tau_n,j})+\log (d_n)-1)(m(\widehat{\gamma}_{\tau_n,\ell})+\log (d_n)-1) &  \\[10pt] 
 +\displaystyle (m(\widehat{\gamma}_{\tau_n,j})+\log d_n)\int_0^1 \widehat{R}_{\tau_n,j,\ell}(u,1)\frac{\diff u}{u} + (m(\widehat{\gamma}_{\tau_n,\ell})+\log d_n)\int_0^1 \widehat{R}_{\tau_n,j,\ell}(1,u)\frac{\diff u}{u}\bigg) & \mbox{if } j<\ell.
\end{cases}
\end{align*}
This yields an $(1-\alpha)-$asymptotic QB confidence region for the extreme expectile $\boldsymbol{\xi}_{\tau'_n}$ as the deformed random ellipsoid
\[
\widehat{\mathcal{E}}_{\tau'_n,\alpha}^{\star} = \bfxihat_{\tau'_n}^{\star} \exp\left( \left[ \bfVhat_n^{\star,\mathrm{QB}}(\bfgamma,\bfR) \right]^{1/2} B_d\left(\bfzero_d,\sqrt{\chi^2_{d,1-\alpha}/n(1-\tau_n)} \, \log d_n \right) \right).
\]
We can also deduce from this confidence region a QB asymptotic $(1-\alpha)-$confidence interval for the $j$th marginal extreme expectile at level $\tau'_n$:
\begin{align*}
\widehat{I}_{\tau'_n,j,\alpha}^{\star}    &= \left[ \widehat{\xi}_{\tau'_n,j}^{\star} \exp\left( - \frac{\log d_n}{\sqrt{n(1-\tau_n)}} \sqrt{\bfVhat_{n,j,j}^{\star,\mathrm{QB}}(\bfgamma,\bfR)} z_{1-\alpha/2}  \right), \right. \\
                                        & \qquad \left. \widehat{\xi}_{\tau'_n,j}^{\star} \exp\left(  \frac{\log d_n}{\sqrt{n(1-\tau_n)}} \sqrt{\bfVhat_{n,j,j}^{\star,\mathrm{QB}}(\bfgamma,\bfR)} z_{1-\alpha/2}  \right) \right].
\end{align*}
This is an adjusted version of the confidence interval based on the so-called indirect estimator in~\cite{daogirstu2018}. We shall compare the relative finite-sample performance of the intervals $\widetilde{I}_{\tau'_n,j,\alpha}^{\star}$ and $\widehat{I}_{\tau'_n,j,\alpha}^{\star}$, and of the regions $\widetilde{\mathcal{E}}_{\tau'_n,\alpha}^{\star}$ and $\widehat{\mathcal{E}}_{\tau'_n,\alpha}^{\star}$, in Section~\ref{sec:simulations}. 

\subsection{Testing the equality of extreme expectiles}\label{sec:Hypo} 

An alternative way of carrying out joint inference about several risk measures is to test their equality. This is relevant to actuarial and financial practice, where risk managers may want to assess the asymptotic dependence between several risk variables, individual stock prices or stock indices, as well as whether certain assets or stocks should be considered riskier than others. 
We show here how our construction of asymptotic confidence regions can be used to design a test of equality of extreme expectiles. We focus here on properly extreme levels since this is the relevant case for extreme risk management.

Consider, for an order $\tau=\tau'_n\to 1$ where $n(1-\tau'_n)\to c\in [0,\infty)$ as $n\to\infty$, the system of hypotheses
$$
\begin{cases}
H_0: & \xi_{\tau'_n,1}=\cdots=\xi_{\tau'_n,d}=\xi_{\tau'_n}, \\
H_1: & \exists(j,\ell) \mbox{ with } j\neq \ell \mbox{ such that } \xi_{\tau'_n,j}\neq \xi_{\tau'_n,\ell}.
\end{cases}
$$
To construct a testing procedure for this problem, we note that we have at our disposal jointly asymptotically Gaussian estimators of the $\xi_{\tau'_n,j}$. Testing the equality of the $\xi_{\tau'_n,j}$ can thus be essentially viewed as testing the equality of the means of a Gaussian random vector. A simple and powerful solution to this problem is given by a likelihood ratio test, which we briefly recall here; more can be found in {\it e.g.}~\cite{sil1970}. Suppose that $\boldsymbol{Z}=(Z_1,\ldots,Z_d)^{\top}$ is a $d-$dimensional Gaussian random vector with mean $\boldsymbol{m}$ and a {\it known}, positive definite covariance matrix $\bfV$. Suppose that it is of interest to consider the nested models problem
\[
\begin{cases}
M_0: & m_{1}=\cdots=m_{d}=m,\\
M_1: & \exists(j,\ell) \mbox{ with } j\neq \ell \mbox{ such that } m_j\neq m_{\ell}.
\end{cases}
\]
%
The (log-likelihood ratio) deviance statistic for testing the validity of 
model $M_0$ is 
\[
\Lambda = (\boldsymbol{Z} - \widehat{m} \boldsymbol{1}_d)^{\top} \bfV^{-1} (\boldsymbol{Z} - \widehat{m} \boldsymbol{1}_d), \ \mbox{ with } \ \widehat{m} = \frac{\boldsymbol{Z}^{\top} \bfV^{-1} \boldsymbol{1}_d}{\boldsymbol{1}_d^{\top} \bfV^{-1} \boldsymbol{1}_d}.
\]
In model $M_0$, the statistic $\Lambda$ has a chi-square distribution with $d-1$ degrees of freedom.

In our case, 
we can set $\boldsymbol{Z}$ to be the LAWS-based extrapolating estimator $\bfxitilde_{\tau'_n}^{\star}$ or the QB extrapolating estimator $\bfxihat_{\tau'_n}^{\star}$. This leads us to two distinct testing procedures. 
\vskip1ex
\noindent
{\bf LAWS-based test} Following the discussion of Section~\ref{sec:CI_ext}, we approximate the distribution of the vector 
$\boldsymbol{Z} = \boldsymbol{Z}_n = \log \bfxitilde_{\tau'_n}^{\star}+\bfbhat^{\mathrm{QB}}/\sqrt{n(1-\tau_n)}$ by a Gaussian distribution with mean 
$
\boldsymbol{m} = \boldsymbol{m}_n = \log \boldsymbol{\xi}_{\tau'_n} 
$ 
and covariance matrix 
\[
\bfVhat = \bfVhat_n = \frac{\log[(1-\tau_n)/(1-\tau'_n)]}{\sqrt{n(1-\tau_n)}}  \bfVhat_n^{\star,\mathrm{LAWS}}(\bfgamma,\bfR)
\]
with the notation of Sections~\ref{sec:CI_inter} and~\ref{sec:CI_ext}. We thus compute the test statistic 
\[
\Lambda = \Lambda_n^{\mathrm{LAWS}} = (\boldsymbol{Z} - \widehat{m} \boldsymbol{1}_d)^{\top} \bfVhat^{-1} (\boldsymbol{Z} - \widehat{m} \boldsymbol{1}_d), \ \mbox{ with } \ \widehat{m} = \frac{\boldsymbol{Z}^{\top} \bfVhat^{-1} \boldsymbol{1}_d}{\boldsymbol{1}_d^{\top} \bfVhat^{-1} \boldsymbol{1}_d}.
\]
We finally define a test with asymptotic type I error $\alpha$ by deciding that if $\Lambda_n^{\mathrm{LAWS}}>\chi^2_{d-1,1-\alpha}$, where $\chi^2_{d-1,1-\alpha}$ is the $(1-\alpha)-$quantile of the chi-square distribution with $d-1$ degrees of freedom, we reject $H_0$; otherwise, accept $H_0$.
\vskip1ex
\noindent
{\bf QB test} Still following Section~\ref{sec:CI_ext}, we approximate the distribution of the vector 
$\boldsymbol{Z} = \boldsymbol{Z}_n = \log \bfxihat_{\tau'_n}^{\star}$ by a Gaussian distribution with mean $\boldsymbol{m} = \boldsymbol{m}_n = \log \boldsymbol{\xi}_{\tau'_n}$ and covariance matrix 
\[
\bfVhat = \bfVhat_n = \frac{\log[(1-\tau_n)/(1-\tau'_n)]}{\sqrt{n(1-\tau_n)}}  \bfVhat_n^{\star,\mathrm{QB}}(\bfgamma,\bfR)
\]
with the notation of Section~\ref{sec:CI_ext}. We thus compute the test statistic 
\[
\Lambda = \Lambda_n^{\mathrm{QB}} = (\boldsymbol{Z} - \widehat{m} \boldsymbol{1}_d)^{\top} \bfVhat^{-1} (\boldsymbol{Z} - \widehat{m} \boldsymbol{1}_d), \ \mbox{ with } \ \widehat{m} = \frac{\boldsymbol{Z}^{\top} \bfVhat^{-1} \boldsymbol{1}_d}{\boldsymbol{1}_d^{\top} \bfVhat^{-1} \boldsymbol{1}_d}.
\]
A test with asymptotic type I error $\alpha$ is defined by rejecting $H_0$ if and only if  $\Lambda_n^{\mathrm{QB}}>\chi^2_{d-1,1-\alpha}$.

Our goal is now to compare the performance of our inference procedures (asymptotic confidence regions and tests) on simulated data in a variety of models, before showcasing our procedures on a sample of real data. 
%
%
\section{Simulation experiments}
\label{sec:simulations} 

Here we study the finite-sample performance of the inferential methodology developed in Section~\ref{sec:CI}. We first assess the quality of inference about marginal extreme expectiles. 
We then study the performance of our joint confidence regions for intermediate and extreme expectiles. 
Finally, we investigate the power of the tests 
for the equality of extreme expectiles. To save space, all Figures and Tables containing our full results are deferred to Appendix~\ref{sec:additional_results}. 

\subsection{Marginal uncertainty about tail expectiles} 
\label{sec:simulations:1D}


%
Here we simulate $M=10{,}000$ samples of $n=1{,}000$ independent observations from 
\begin{itemize}
\item The Fr\'echet distribution, having distribution function $F(x)=\exp(-x^{-1/\gamma})$ for $x>0$,    
\item The Pareto distribution, having distribution function $F(x)=1-x^{-1/\gamma}$ for $x>1$, 
\item The Student-$t$ distribution with $1/\gamma$ degrees of freedom.
\end{itemize}
The tail index is chosen to be $\gamma=1/3$ in each case. For each simulated sample we estimate the (univariate) expectile at the extreme level $\tau_n'=0.999=1-1/n$ and we compute the associated confidence intervals $\widetilde{I}_{\tau'_n,\alpha}^{\star}$ and $\widehat{I}_{\tau'_n,\alpha}^{\star}$ defined in Section~\ref{sec:CI_ext} (there is no dependence on the label of the marginal in this univariate case) with $95\%$ nominal coverage probability. The anchor intermediate level is taken to be $\tau_n=1-k/n$, with $k\in [6,300]$. 
Then, we compute a Monte Carlo approximation of the relative Mean Squared Error (MSE) for the extrapolating point estimators and the actual coverage probability of the corresponding interval estimators. Results are collected in Figure~\ref{fig:univariate_extreme_level}, see Appendix~\ref{sec:additional_results}. 

The top panels of this Figure show that the QB extrapolating estimator has lower relative MSE than its LAWS counterpart in the Fr\'echet and Pareto cases, and comparable MSE in the Student-$t$ case. Interestingly, however, the adjusted interval estimators perform comparably in each case, and in fact the LAWS confidence interval has slightly better and more stable coverage, as the middle and bottom panels show. Our adjusted intervals provide visibly improved results compared to their unadjusted versions for all three distributions, with a remarkable improvement in the LAWS case for the Fr\'echet and Pareto distributions. By contrast, the actual non-coverage probability of the unadjusted versions is typically in the range of $15$-$25\%$. As a conclusion, it appears that in terms of marginal inference at the extreme level, the LAWS and QB extrapolating estimators are comparable, with a slight advantage for the former once our adjustment to the confidence interval has been applied. 

\subsection{Joint inference about intermediate and extreme expectiles}
\label{sec:simulations:joint}

In the second and third parts of our experiments we work with, among others, two families of Archimedean copulae, which we briefly introduce below. Further details can be found in~\citet{joe2014}. Let $\varphi: (0,1] \to [0,\infty)$ be a convex and strictly decreasing function with $\varphi(1)=0$ and $\varphi(t)\uparrow \infty$ as $t\downarrow 0$. The Archimedean copula in dimension $d$ with generator $\varphi$ is the $d$-dimensional distribution function $C$ with uniform marginals defined by
\[
C(\bfu)= \varphi^{-1}(\varphi(u_1)+\cdots+\varphi(u_d)), \ \bfu=(u_1,\ldots,u_d)\in[0,1]^d.
\]
The Archimedean families we consider are, first, the Clayton family, defined through the generator $\varphi(u)=\theta^{-1}(u^{-\theta}-1)$ for $\theta>0$. Here the components of $\bfu$ become independent for $\theta\to 0$, and completely dependent for $\theta\to\infty$. We also consider the Gumbel family, defined through the generator $\varphi(u)=(-\log(u))^{\vartheta}$ for $\vartheta\geq 1$, with $\vartheta=1$ representing the case of independent variables and $\vartheta\to\infty$ the case of perfectly dependent variables.

Our experiments are based on the below models for $\bfX=(X_1,\ldots,X_d)$ (we take $d\leq 5$). 
\begin{enumerate}[label=(\roman*)]
\item {[Clayton-Fr\'echet model]} Let $\bfU$ follow a Clayton copula with dependence parameter $\theta=10$. Take $X_j=(-\log(U_j))^{-\gamma}$ with $\gamma=1/3$. Then $\bfX$ has Fr\'echet marginal distributions with tail index $1/3$ and a Clayton copula dependence structure.
\item {[Gaussian-Student model]} Let $\bfU$ follow a Gaussian copula. Pairwise correlation parameters are taken as $\rho_{1,2}=0.8$ for $d=2$, $(\rho_{1,2}=0.8,\,\rho_{1,3}=0.6,\,\rho_{2,3}=0.4)$ for $d=3$, $(\rho_{1,2}=0.8,\,\rho_{1,3}=0.6,\,\rho_{1,4}=0.4,\, \rho_{2,3}=0.5,\, \rho_{2,4}=0.4,\, \rho_{3,4}=0.4)$ for $d=4$ and $(\rho_{1,2}=0.8,\,\rho_{1,3}=0.6,\,\rho_{1,4}=0.4,\, \rho_{1,5}=0.2,\, \rho_{2,3}=0.5,\, \rho_{2,4}=0.4,\, \rho_{2,5}=0.3,\, \rho_{3,4}=0.6,\, \rho_{3,5}=0.4,\, \rho_{4,5}=0.3)$ for $d=5$. Take $X_j= F_\nu^{-1}(U_j)$ where $F_{\nu}$ is the Student-$t$ distribution function with $\nu=3$ degrees of freedom. Then $\bfX$ has Student-$t$ marginal distributions with tail index $1/3$ and a Gaussian copula dependence structure.
\item {[Gumbel-Fr\'echet model]} Let $\bfU$ follow a Gumbel copula with dependence parameter $\vartheta=3$. Take $X_j=(-\log(U_j))^{-\gamma}$ with $\gamma=1/3$. Then $\bfX$ has Fr\'echet marginal distributions with tail index $1/3$ and a Gumbel copula dependence structure.
\item {[Multivariate Student-$t$ model]} Let $\bfX$ follow a zero-mean multivariate Student-$t$ distribution with $\nu=3$ degrees of freedom and a 
scale matrix given by $\rho_{1,2}=0.8$ for $d=2$, $(\rho_{1,2}=0.8,\,\rho_{1,3}=0.6,\,\rho_{2,3}=0.4)$ for $d=3$, $(\rho_{1,2}=0.8,\,\rho_{1,3}=0.6,\,\rho_{1,4}=0.4,\, \rho_{2,3}=0.5,\, \rho_{2,4}=0.4,\, \rho_{3,4}=0.4)$ for $d=4$ and $(\rho_{1,2}=0.8,\,\rho_{1,3}=0.6,\,\rho_{1,4}=0.4,\, \rho_{1,5}=0.2,\, \rho_{2,3}=0.5,\, \rho_{2,4}=0.4,\, \rho_{2,5}=0.3,\, \rho_{3,4}=0.6,\, \rho_{3,5}=0.4,\, \rho_{4,5}=0.3)$ for $d=5$.
\end{enumerate}
In these four models, all univariate margins have the same tail index $\gamma=1/3$. The components of $\bfX$ are asymptotically independent in models (i) and (ii), in the sense that all pairwise tail copulae are identically 0, and asymptotically dependent in models (iii) and (iv). 
%
%
Figure~\ref{fig:examples} in Appendix~\ref{sec:additional_results} shows typical samples from each model. It is important to note that even though models (i) and (ii) are technically cases of tail independence, finite samples can show a degree of dependence in the joint empirical tail. 
We also highlight that a sample generated from models (iii) or (iv) typically shows strong dependence in the joint upper tail. 

We first study the finite-sample behaviour of the intermediate expectile estimators. In each model, we simulate $M=10^4$ samples of size $n=m\cdot 10^3$, with $m\in \{ 1,1.5,2,2.5,5,10\}$ and dimension $d\in\{2,3,4,5\}$. We estimate the $d$-dimensional expectile $\boldsymbol{\xi}_{\tau_n}$, with $\tau_n=1-1/\sqrt{n}$, 
using the LAWS and QB expectile point estimators and the confidence regions $\widetilde{\mathcal{E}}_{\tau_n,\alpha}$ and $\widehat{\mathcal{E}}_{\tau_n,\alpha}$, with $\alpha=0.05$ ($95\%$ nominal coverage probability), described in Section~\ref{sec:CI_inter}. 
Then, we compute a Monte Carlo approximation of the relative MSE of the LAWS and QB point estimators across all components 
and we report the actual (non-)coverage probabilities of the associated confidence regions (see Tables~\ref{tab:RMSE_expect_LAWS_QB},~\ref{tab:cov_prob_expc_int_laws} and~\ref{tab:cov_prob_expc_int_QB} in Appendix~\ref{sec:additional_results}).

With every model except the Gumbel-Fr\'echet model, the actual coverage probability of the LAWS confidence region estimator is close to the nominal level. With the Gumbel-Fr\'echet model, permissive confidence regions are generally obtained. This seems to be due to the strong dependence structure of the Gumbel-Fr\'echet model which is somewhat difficult to estimate accurately. 
The conclusions for the QB confidence region are similar. 
By contrast, the naive confidence regions obtained assuming that the margins are independent (and thus ignoring the question of the estimation of the asymptotic dependence between components) provide unsuitable regions whose actual non-coverage probabilities are either substantially higher than desired (for the LAWS estimator) or virtually equal to zero (for the QB estimator). Our proposal therefore allows to obtain considerably more accurate confidence regions than existing methods; moreover, while the LAWS confidence region performs best in the Clayton-Fr\'{e}chet and Gaussian-Student-$t$ models, the QB confidence region is better in the Gumbel-Fr\'echet and Multivariate Student-$t$ models, and results do not seem to deteriorate significantly with increasing dimension (at least up to $d=5$).

To assess the performance of our methods at the extreme level, we keep the same simulation setting but with the difference that a single sample size $n=1{,}000$ and the extreme level $\tau_n'=0.999=1-1/n$ are used. Monte Carlo approximations of 
the actual coverage probabilities are displayed in 
Figure~\ref{fig:cov_prob_laws_QB_extreme}, 
see Appendix~\ref{sec:additional_results}. 
Our proposed confidence region estimators provide satisfactory estimation results at the extreme level, with the exception of the QB region in the asymptotically independent case of the Clayton-Fr\'echet model. The LAWS-based confidence region seems to perform well, with very stable coverage probabilities close to the nominal level in Fr\'echet models, and a clearly identified stability region for values of $k$ around 50 with a coverage probability close to the nominal level in Student models. There is no clear conclusion as to which method is best in a given case, with the LAWS method being at times slightly more conservative than the QB method, and in other models slightly more permissive. Results seem to be robust with respect to the dimension. 


\subsection{Testing the equality of extreme expectiles}
\label{sec:simulations:testing}

In our final simulation experiment we check the performance of the tests for equality of several extreme expectiles. We keep the models of Section~\ref{sec:simulations:joint}, although in each of the models (i)-(iv) we allow the tail index $\gamma$ to vary within the interval $[0.1, 0.4]$, for one margin of the joint distribution. In each case we simulate $M=10{,}000$ samples of size $n=1{,}000$ from the thus modified models. The null hypothesis of equal extreme expectiles, {\it i.e.}~$H_0:\xi_{\tau_n',1}=\cdots=\xi_{\tau_n',d}=\xi_{\tau_n'}$ is then true if and only if $\gamma=1/3$. Then we perform the LAWS and QB tests and we compute the proportion of rejections, thus deriving a Monte Carlo approximation of the type I error probability and the corresponding power of the test.

Table~\ref{tab:significance_test} in Appendix~\ref{sec:additional_results} reports the type I errors of the LAWS and QB versions of the test for $\tau_n'=0.999=1-1/n$, $k=50$ and $d=2,3,4,5$. The QB version has a larger type I error than anticipated in the case of the Clayton-Fr\'echet model; in the other cases, our tests tend to have a lower type I error than expected. However, results obtained with the LAWS version tend to improve as the dimension increases, approaching the nominal level when $d=5$. Figure~\ref{fig:power_test} in Appendix~\ref{sec:additional_results} displays the power of both versions of the test when $\gamma\in [0.1, 0.4]$ and $d=2,3,4,5$. The power curves reflect the excellent power of both tests. The rejection rate increases (decreases) for stronger (weaker) dependence structures and the highest (lowest) rejection rate is indeed obtained with the Gumbel-Fr\'echet (Clayton-Fr\'echet) model. Our testing procedures appear to yield reasonably stable results across a wide range of parameters $k$, as Figure~\ref{fig:cov_prob_laws_extreme} in Appendix~\ref{sec:additional_results} shows in the case $d=2$.

%
%
\section{Risk analysis of multiple exchange rates}\label{sec:real_data} 
%

The analysis of exchange rate risk is one of the most difficult tasks in economics. Links between exchange rates and fundamental economic principles have been established \citep[see {\it e.g.}][]{engwes2005}. A modern approach to understanding exchange rates 
uses a supply-and-demand analysis of the exchange rate seen as the price of domestic assets in terms of foreign assets \citep[see][]{madura2014financial}. The exchange rate is influenced by a positive interest rate differential, in the short term, implying an appreciation of the home currency. In the long term, all other things being equal, a rise in a country's price level is correlated with depreciation of its currency, while an increased demand for exports (imports) is correlated with appreciation (depreciation) of its currency, see {\it e.g.}~\citet[][p.~201]{harsmidav1992}.

We consider negative weekly log-returns (returns for brevity) of the exchange rates of the Great British Pound (GBP) versus the United States Dollar (USD), the Japanese Yen (JPY), the Canadian Dollar (CAD), the Australian Dollar (AUD) and the Norwegian Krone (NOK), from January 1, 1980 to June 26, 2020\footnote{Available from {\tt https://www.investing.com/}.}. 
These samples of size $n=2{,}133$ are plotted on the top panels of Figure \ref{fig:ExcRatetailIndex} in Appendix \ref{sec:additional_results}. They are technically, of course, time series data; in our results we do not enter into the important but difficult question of handling serial dependence. This is the reason why, as suggested by~\cite{caieinhaazho2015}, we chose to consider weekly returns as a way to substantially reduce the amount of dependence present in the exchange rates.
The United States and Japan are developed, industrialised economies characterised by the presence of a large number of global firms and, in recent years, similar monetary policy leading to low interest rates, therefore a substantial degree of dependence between the GBP-USD and GBP-JPY exchange rates is to be expected. Canada and Australia are close partners of United States, accessing the American market for exports, attracting American capital and technology for economic development and sharing large international finance institutions. Hence, a fairly strong dependence among the GBP-USD, GBP-CAD and GBP-AUD exchange rates is expected as well. Such expectations are confirmed from the scatterplots in Figure~\ref{fig:PredictedRiskRegions} (see also Figure \ref{fig:PredictedConfidenceRegionsWeakDep} in Appendix \ref{sec:additional_results}). We also find visible dependence within the (GBP-CAD, GBP-NOK) and (GBP-AUD, GBP-NOK) pairs. Table~\ref{tab:real_data_dependence} gives estimated correlations between exchange rates, suggesting strong correlations between GBP-USD and GBP-JPY, GBP-USD and GBP-CAD. 

The purpose of analysing multiple exchange rate returns simultaneously is that it can be useful in understanding and predicting the risks that nations and companies exposed to the global economy are subjected to. Risk analysis is most often based on Value-at-Risk (VaR) at the $99.9\%$ level~\citep[see {\it e.g.}][]{dre2003, haamerzho2016} or on a quantile at level $1-p_n$ where $p_n$ is not larger than $1/n$. The potential of extreme expectiles for risk assessment is illustrated by~\cite{beldib2017},~\cite{daogirstu2018} and~\cite{padstu2020}, where it is found that parametric and nonparametric expectile-based forecasts may provide similar outcomes to those obtained with VaR, in suitable settings. We analyse here the joint tail risk in multiple exchange rate returns through our expectile-based multivariate inferential procedures, at the extreme level $\tau_n'=1-p_n=0.9995312$ with $p_n=1/n$.

%
%
%
\begin{table}[b!]
\centering
\begin{tabular}{cccccc}
\toprule
& GBP-USD & GBP-JPY & GBP-CAD & GBP-AUD & GBP-NOK\\
GBP-USD & -- & 0.559 & 0.751 & 0.481 & 0.337\\
GBP-JPY & 1.667 & -- & 0.389 & 0.273 & 0.361 \\
GBP-CAD & 1.460 & 1.753 & -- & 0.646 & 0.438\\
GBP-AUD & 1.613 & 1.753 & 1.533 & -- & 0.405\\
GBP-NOK & 1.720 & 1.760 & 1.686 & 1.673 & --\\
\bottomrule
\end{tabular}
\caption{Pairwise empirical correlations (upper off-diagonal values) and extremal coefficient estimates (lower off-diagonal values) of exchange rate returns.}
\label{tab:real_data_dependence}
\end{table}
%

Point estimates and the $95\%$ confidence intervals of the tail index for the five series are displayed in the middle row of Figure \ref{fig:ExcRatetailIndex} in Appendix \ref{sec:additional_results}. The tails of the individual series seem moderately heavy; estimates are fairly stable for a series-dependent interval of values of $k$. To select a common range, we plot the trace of the estimated variance-covariance matrix $\bfVhat_n^{\star,\mathrm{LAWS}}(\bfgamma,\bfR)$ relative to the extrapolating estimator $\bfxitilde^{\star}_{\tau_n'}$ (discussed in Section \ref{sec:CI_ext}) that combines together individual information coming from the five exchange rates returns.
Figure \ref{fig:KSelectionTraceVar} suggests that the trace of $\bfVhat_n^{\star,\mathrm{LAWS}}(\bfgamma,\bfR)$ is stable for $k\in[50,150]$. 
In the sequel, we use $k=150$ in our inferential procedures. Tail index point estimates of individual exchange rate returns with corresponding $95\%$ confidence intervals are reported in Table~\ref{tab:final_estimate_results}.
%
%
\begin{table}[t!]
\centering
\begin{tabular}{lccc}
\toprule
& \multicolumn{3}{c}{Estimator}\\
\cdashlinelr{2-4}
Exchange rate & $\widehat{\gamma}_n$  &  $\widetilde{\xi}^{\star}_{\tau_n'}$ & $\widehat{\xi}^{\star}_{\tau_n'}$ \\
\midrule
GBP-USD & $0.3331 \, [0.2798,\,0.3865]$ & $0.0716 \, [0.0517,\,0.0991]$
& $0.0774 \, [0.0565,\,0.1060]$\\[5pt]
GBP-JPY & $0.3822 \, [0.3210,\,0.4433]$ & $0.1212 \, [0.0815,\,0.1786]$
& $0.1319 \, [0.0901,\,0.1931]$\\[5pt]
GBP-CAD & $0.3365 \, [0.2826,\,0.3903]$ & $0.0719 \, [0.0521,\,0.1005]$
& $0.0788 \, [0.0572,\,0.1083]$\\[5pt]
GBP-AUD & $0.3628 \, [0.3047,\,0.4208]$ & $0.0927 \, [0.0649,\,0.1343]$
& $0.1019 \, [0.0715,\,0.1451]$\\[5pt]
GBP-NOK & $0.3360 \, [0.2823,\,0.3898]$ & $0.0624 \, [0.0456,\,0.0880]$
& $0.0700 \, [0.0509,\,0.0962]$\\
\bottomrule
\end{tabular}
\caption{Tail index and extreme expectile estimates relative to the exchange rate returns, obtained with $k=150$ and $\tau_n'=0.9995312$. Between square brackets are $95\%$ confidence intervals.}
\label{tab:final_estimate_results}
\end{table}
%
%

The lower off-diagonal values in Table \ref{tab:real_data_dependence} are pairwise extremal coefficient estimates. Recall that the bivariate extremal coefficient is a tail dependence measure $\omega\in [1, 2]$, equal to the value at $(1,1)$ of the stable tail dependence function~\citep{drehua1998}, with the lower and upper bounds representing the case of complete dependence and independence~\citep[see {\it e.g.}][]{berpad2015}. For two exchange rates labelled $j$ and $\ell$, say, their extremal coefficient is estimated with $\widehat{\omega}_{n,j,\ell}=2-\widehat{R}_{\tau_n,j,\ell}(1,1)$, where $\widehat{R}_{\tau_n,j,\ell}(1,1)$ is defined in~\eqref{eqn:tail_copula_est}. These suggest that there is a fairly strong dependence in the joint tail of the two-dimensional exchange rate returns (GBP-USD, GBP-CAD) and (GBP-CAD, GBP-AUD), with milder dependence in the other pairs of returns. 
%
%
\begin{table}[t!]
\centering
\begin{tabular}{llllc}
\toprule
Null Hypothesis (for $\Lambda_n^{\mathrm{LAWS}}$ and $\Lambda_n^{\mathrm{QB}}$) & $\Lambda_n^{\mathrm{LAWS}}$  &  $\Lambda_n^{\mathrm{QB}}$ &  $\Lambda_n^{\mathrm{Q}}$ & $\chi^2_{d-1,1-\alpha}$ \\
\midrule
$\xi_{\tau_n',\mathrm{GBP-USD}}=\xi_{\tau_n',\mathrm{GBP-JPY}}=\xi_{\tau_n',\mathrm{GBP-CAD}}$ & &  & \\[5pt]
$\;\qquad\qquad\quad=\xi_{\tau_n',\mathrm{GBP-AUD}}=\xi_{\tau_n',\mathrm{GBP-NOK}}$ & $9.8919^{(*)}$ & $12.2167^{(*)}$ & 5.9914 & $9.4877$\\[5pt]
\cdashlinelr{1-5}
$\xi_{\tau_n',\mathrm{GBP-USD}}=\xi_{\tau_n',\mathrm{GBP-CAD}}=\xi_{\tau_n',\mathrm{GBP-AUD}}$ &  &  & \\[5pt]
$\;\qquad\qquad\qquad\qquad\qquad\quad\ \ =\xi_{\tau_n',\mathrm{GBP-NOK}}$ & $3.3015$ & $4.2765$ & 2.0412 &$7.8147$\\[5pt]
\cdashlinelr{1-5}
$\xi_{\tau_n',\mathrm{GBP-USD}}=\xi_{\tau_n',\mathrm{GBP-JPY}}$ 
& $5.4383^{(*)}$ & $7.1357^{(*)}$ & 3.3527 & $3.8415$\\[5pt]
$\xi_{\tau_n',\mathrm{GBP-USD}}=\xi_{\tau_n',\mathrm{GBP-CAD}}$ 
& $0.0033$ & $0.0143$ & 0.0052 & --\\[5pt]
$\xi_{\tau_n',\mathrm{GBP-USD}}=\xi_{\tau_n',\mathrm{GBP-AUD}}$ 
& $1.5692$ & $2.2084$ & 1.0083 & --\\[5pt]
$\xi_{\tau_n',\mathrm{GBP-USD}}=\xi_{\tau_n',\mathrm{GBP-NOK}}$ 
& $0.3410$ & $0.2895$ & 0.1610 & --\\[5pt]
$\xi_{\tau_n',\mathrm{GBP-JPY}}=\xi_{\tau_n',\mathrm{GBP-CAD}}$ 
& $4.6414^{(*)}$ & $5.6585^{(*)}$ & 2.7691 & --\\[5pt]
$\xi_{\tau_n',\mathrm{GBP-JPY}}=\xi_{\tau_n',\mathrm{GBP-AUD}}$ 
& $1.0525$ & $1.3209$ & 0.6611 & --\\[5pt]
$\xi_{\tau_n',\mathrm{GBP-JPY}}=\xi_{\tau_n',\mathrm{GBP-NOK}}$ 
& $7.3940^{(*)}$ & $8.5705^{(*)}$ & $4.2973^{(*)}$ & --\\[5pt]
$\xi_{\tau_n',\mathrm{GBP-CAD}}=\xi_{\tau_n',\mathrm{GBP-AUD}}$ 
& $1.5539$ & $2.3027$ & 1.0146 & --\\[5pt]
$\xi_{\tau_n',\mathrm{GBP-CAD}}=\xi_{\tau_n',\mathrm{GBP-NOK}}$ 
& $0.4111$ & $0.4199$ & 0.2191 & --\\[5pt]
$\xi_{\tau_n',\mathrm{GBP-AUD}}=\xi_{\tau_n',\mathrm{GBP-NOK}}$ 
& $3.1154$ & $3.8023$ & 1.8272 & --\\[5pt]
\bottomrule
\end{tabular}
\caption{Hypothesis testing outcome for the exchange rate returns data, obtained with $k=150$ and $\tau_n'=0.9995312$. Starred test statistics indicate rejection at the $\alpha=5\%$ significance level. The null hypothesis tested by $\Lambda_n^{\mathrm{Q}}$ is obtained by replacing expectiles in the first column by their quantile counterparts.}
\label{tab:hypo_testing_results}
\end{table}
%
%
In addition to tail index estimates, Table~\ref{tab:final_estimate_results} reports the expectile point estimates obtained with the extrapolating LAWS estimator $\bfxitilde^{\star}_{\tau_n'}$ and QB estimator $\bfxihat^{\star}_{\tau_n'}$ with associated marginal confidence intervals $\widetilde{I}_{\tau'_n,j,\alpha}^{\star}$ and $\widehat{I}_{\tau'_n,j,\alpha}^{\star}$. We have also computed the two- and three-dimensional asymptotic $95\%$ confidence regions for all the pairs and triplets of exchange rate returns, using the LAWS and QB confidence region estimators $\widetilde{\mathcal{E}}_{\tau'_n,\alpha}^{\star}$ and $\widehat{\mathcal{E}}_{\tau'_n,\alpha}
^{\star}$. Figure~\ref{fig:PredictedRiskRegions} displays these estimated regions for the most tail dependent pairs and triplets of exchange rate returns (plots for other pairs and triplets are available in Figure~\ref{fig:PredictedConfidenceRegionsWeakDep}, see Appendix~\ref{sec:additional_results}). These devices are an important tool for the quantification of the potential contamination risk that a certain type of international economy might be subjected to, and therefore could be useful for risk managers.

Finally, we complete the analysis by performing our testing procedures to assess the validity of the assumption of equal risk severity among exchange rate returns. We did this applying the two versions of the test described in Section~\ref{sec:Hypo}. The hypothesis of equal expectile risk severity among all exchange rate returns is rejected with $5\%$ significance level using both versions of the test (see Table~\ref{tab:hypo_testing_results}). Then, we perform the tests again, assuming the same expectile risk severity between pairs of exchange rate returns only. The outcome of the pairwise tests suggest to reject the null hypothesis with $5\%$ significance level for pairs involving the GBP-JPY exchange rate (except for the (GBP-JPY, GBP-AUD) pair). This suggests that overall the GBP-JPY exchange rate return seems to carry different extreme risk than the other returns; it is interesting to note that this is not obvious either from marginal tail index confidence intervals or extreme expectile confidence intervals, which strongly overlap across marginals. Leaving out the GBP-JPY exchange rate and testing again for equality of extreme expectiles does not give empirical evidence to reject the null hypothesis, confirming our intuition. By way of comparison, we carried out an analogue test on the equality of extreme quantiles, which is built on the joint asymptotic normality of the Weissman quantile estimators across marginals: 
\begin{align*}
 & \frac{\sqrt{n(1-\tau_n)}}{\log[(1-\tau_n)/(1-\tau_n')]}\left(\frac{\widehat{q}_{\tau_n',j}^{\star}}{q_{\tau'_n,j}}-1\right)_{1\leq j\leq d} \tod \mathcal{N}_{d}\left(\bfb^{\star}, \bfV^{\star}(\bfgamma,\bfR) \right), \\
 & \mbox{with } \widehat{q}_{\tau_n',j}^{\star}=\left(\frac{1-\tau_n'}{1-\tau_n}\right)^{-\widehat{\gamma}_{\tau_n,j}}X_{n-\floor*{n(1-\tau_n)},n,j}.
\end{align*}
(The proof is identical to that of Theorem~\ref{theo:asyext}). Neglecting the bias term, and setting $\boldsymbol{Z} = \boldsymbol{Z}_n = \log \bfqhat_{\tau'_n}^{\star}$ as well as
\[
\bfVhat = \bfVhat_n = \frac{\log[(1-\tau_n)/(1-\tau'_n)]}{\sqrt{n(1-\tau_n)}} \times \begin{cases} \widehat{\gamma}_{\tau_n,j}^2 & \mbox{if } j=\ell, \\ \widehat{\gamma}_{\tau_n,j}\widehat{\gamma}_{\tau_n,\ell} \widehat{R}_{\tau_n,j,\ell}(1,1) & \mbox{if } j<\ell, \end{cases}
\]
we then consider the test statistic 
\[
\Lambda = \Lambda_n^{\mathrm{Q}} = (\boldsymbol{Z} - \widehat{m} \boldsymbol{1}_d)^{\top} \bfVhat^{-1} (\boldsymbol{Z} - \widehat{m} \boldsymbol{1}_d), \ \mbox{ with } \ \widehat{m} = \frac{\boldsymbol{Z}^{\top} \bfVhat^{-1} \boldsymbol{1}_d}{\boldsymbol{1}_d^{\top} \bfVhat^{-1} \boldsymbol{1}_d}
\]
in order to test the hypothesis $H_0$~: $q_{\tau'_n,1}=\cdots=q_{\tau'_n,d}=q_{\tau'_n}$. When $\Lambda_n^{\mathrm{Q}}>\chi^2_{d-1,1-\alpha}$, the test rejects this hypothesis with asymptotic type I error $\alpha$. Table~\ref{tab:hypo_testing_results} reports the results of the test applied to exchange rate returns data.
It is readily seen here that this test is much less conclusive than our expectile-based tests, with only the hypothesis $q_{\tau_n',\mathrm{GBP-JPY}}=q_{\tau_n',\mathrm{GBP-NOK}}$ being narrowly rejected. As a result our inferential methodology based on the expectile risk measure appears to be more sensitive than its quantile-based competitor in detecting differences in tail risk, suggesting that the use of expectile-based inference is beneficial in tail risk assessment.

%
%
\begin{figure}[t!]
\centering
\includegraphics[width=0.3\textwidth, page=1]{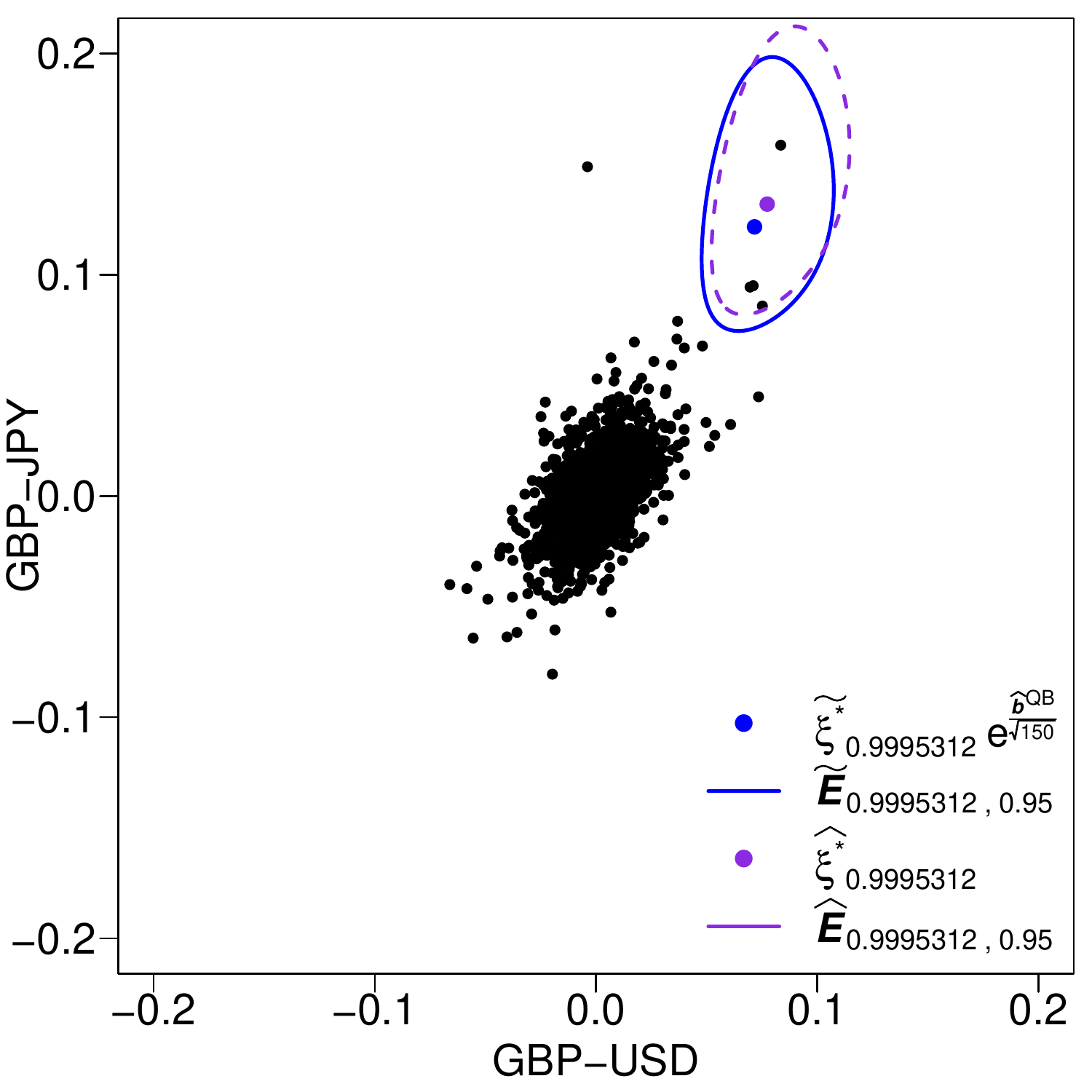}
\includegraphics[width=0.3\textwidth, page=2]{ConfReg2dReturnsExchangeRates.pdf}
\includegraphics[width=0.3\textwidth, page=3]{ConfReg2dReturnsExchangeRates.pdf}\\
\includegraphics[width=0.3\textwidth, page=8]{ConfReg2dReturnsExchangeRates.pdf}
\includegraphics[width=0.3\textwidth, page=9]{ConfReg2dReturnsExchangeRates.pdf}
\includegraphics[width=0.3\textwidth, page=10]{ConfReg2dReturnsExchangeRates.pdf}\\
\includegraphics[width=0.3\textwidth, page=1]{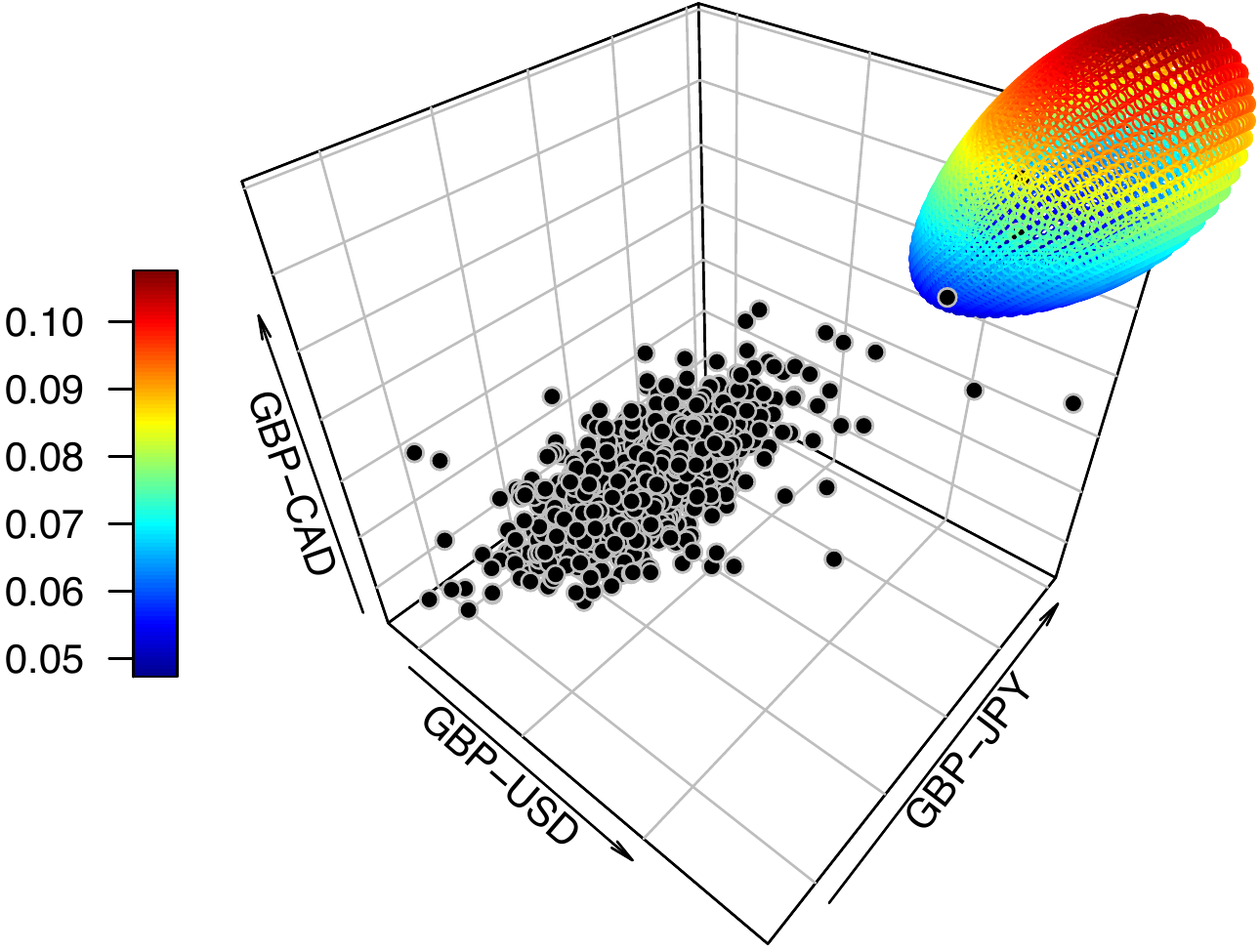}
\includegraphics[width=0.3\textwidth, page=2]{ConfReg3dReturnsEchangeRates.pdf}
\includegraphics[width=0.3\textwidth, page=3]{ConfReg3dReturnsEchangeRates.pdf}\\
\includegraphics[width=0.3\textwidth, page=4]{ConfReg3dReturnsEchangeRates.pdf}
\includegraphics[width=0.3\textwidth, page=6]{ConfReg3dReturnsEchangeRates.pdf}
\includegraphics[width=0.3\textwidth, page=10]{ConfReg3dReturnsEchangeRates.pdf}
\caption{Two- and three-dimensional $95\%$ confidence regions estimates for some pairs and triplets of exchange rate returns, obtained with $k=150$ and $\tau_n'=0.9995312$. In the three-dimensional case we only report LAWS-based confidence regions.}
\label{fig:PredictedRiskRegions}
\end{figure}
%
%
%
%


\section*{Acknowledgements}
\noindent
Part of this research was carried out when the authors were visiting each other at Bocconi University and the University of Nottingham where G. Stupfler was previously based. Support from the Nottingham PEF Fund is gratefully acknowledged. S.A.\! Padoan is supported by the Bocconi Institute for Data Science and Analytics (BIDSA), Italy. G. Stupfler is supported by the French National Research Agency under the grant ANR-19-CE40-0013-01/ExtremReg project.
\bibliographystyle{imsart-nameyear}

\clearpage

\appendix
\renewcommand\thefigure{\Roman{figure}}  
\renewcommand\thetable{\Roman{table}}  
\renewcommand\theequation{\thesection.\arabic{equation}}  
\setcounter{figure}{0}    
\setcounter{table}{0}
\setcounter{equation}{0}

\section{Additional finite-sample results}\label{sec:additional_results}

\begin{figure}[t!]
\centering
\includegraphics[width=0.32\textwidth, page=1]{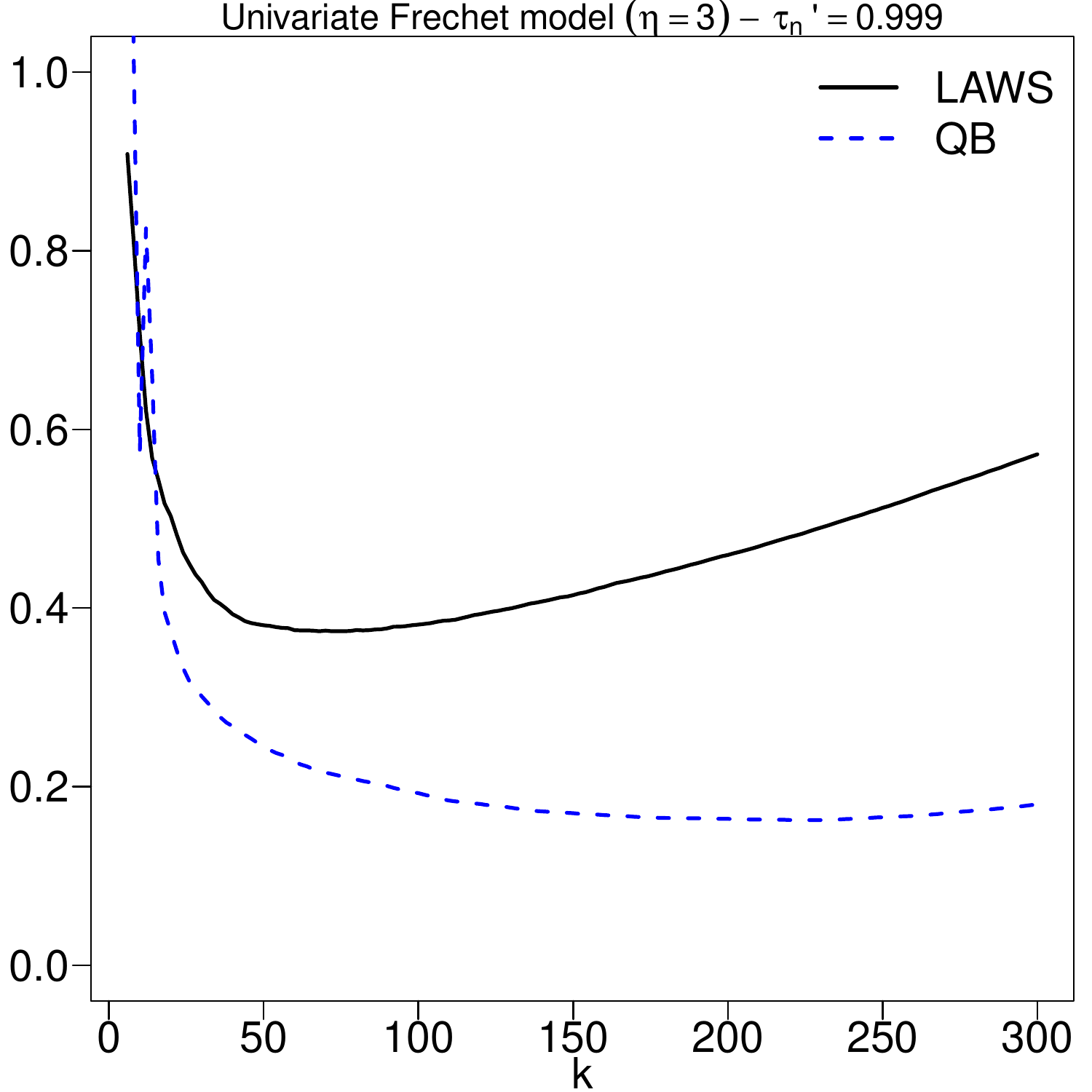}
\includegraphics[width=0.32\textwidth, page=2]{MSEUnivariate.pdf}
\includegraphics[width=0.32\textwidth, page=3]{MSEUnivariate.pdf}\\

\includegraphics[width=0.32\textwidth, page=1]{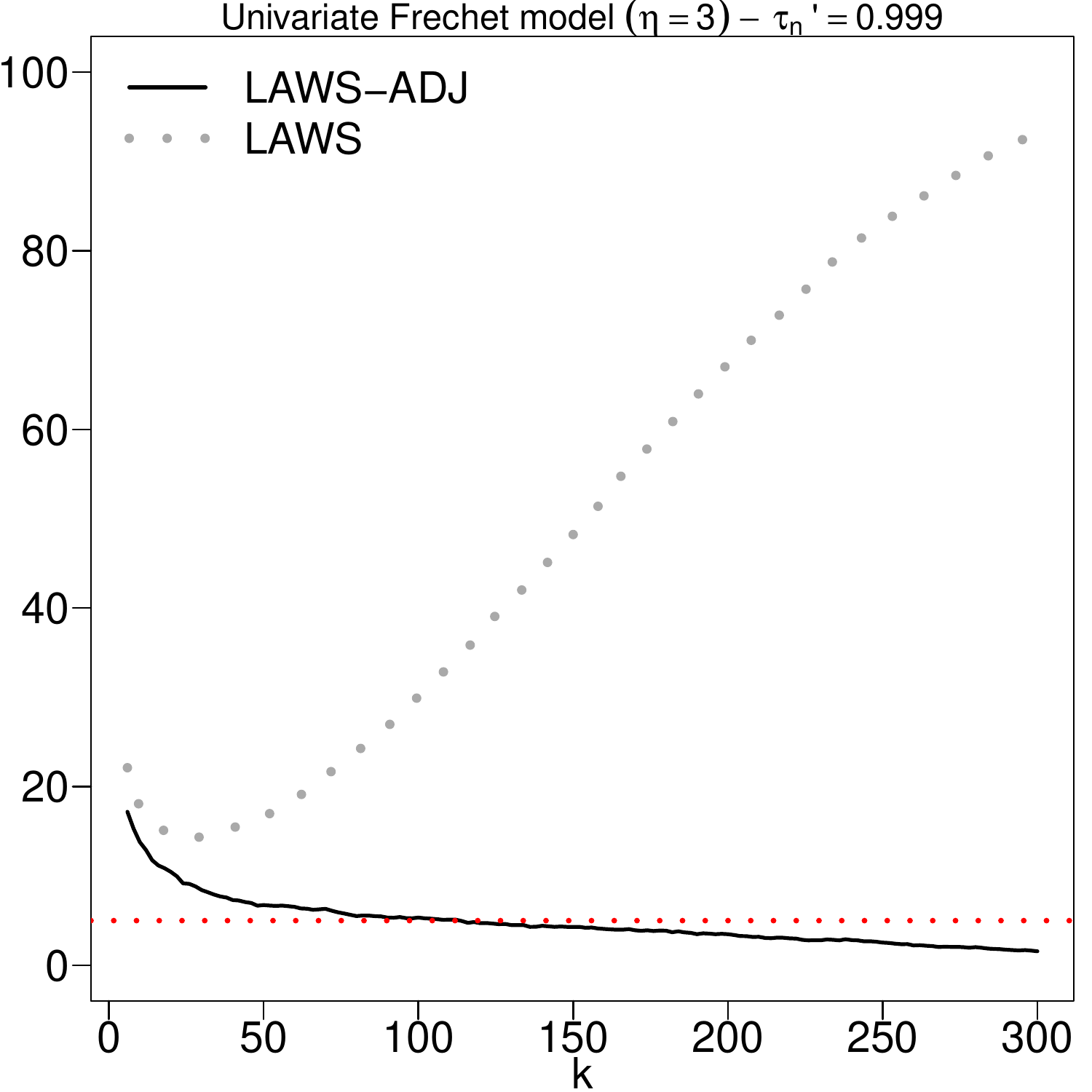}
\includegraphics[width=0.32\textwidth, page=3]{CovProbUnivariate.pdf}
\includegraphics[width=0.32\textwidth, page=5]{CovProbUnivariate.pdf}\\
\includegraphics[width=0.32\textwidth, page=2]{CovProbUnivariate.pdf}
\includegraphics[width=0.32\textwidth, page=4]{CovProbUnivariate.pdf}
\includegraphics[width=0.32\textwidth, page=6]{CovProbUnivariate.pdf}
\caption{Top row: Monte Carlo relative MSE (reported in $\sqrt{\cdot}$) of the univariate versions of the extrapolating LAWS and QB estimators.  
Middle and bottom row: actual non-coverage probabilities (in \%) of the confidence intervals of $\xi_{\tau_n'}$, with $n=1{,}000$, $\tau_n'=0.999$ and $95\%$ nominal level (middle: LAWS method, bottom: QB method), where in each case the dotted line stands for the naive interval and the solid line for its adjusted counterpart. The horizontal dotted red line represents the $5\%$ nominal non-coverage probability.}
\label{fig:univariate_extreme_level}
\end{figure}

\begin{figure}[b!]
\centering
\includegraphics[width=0.45\textwidth, page=1]{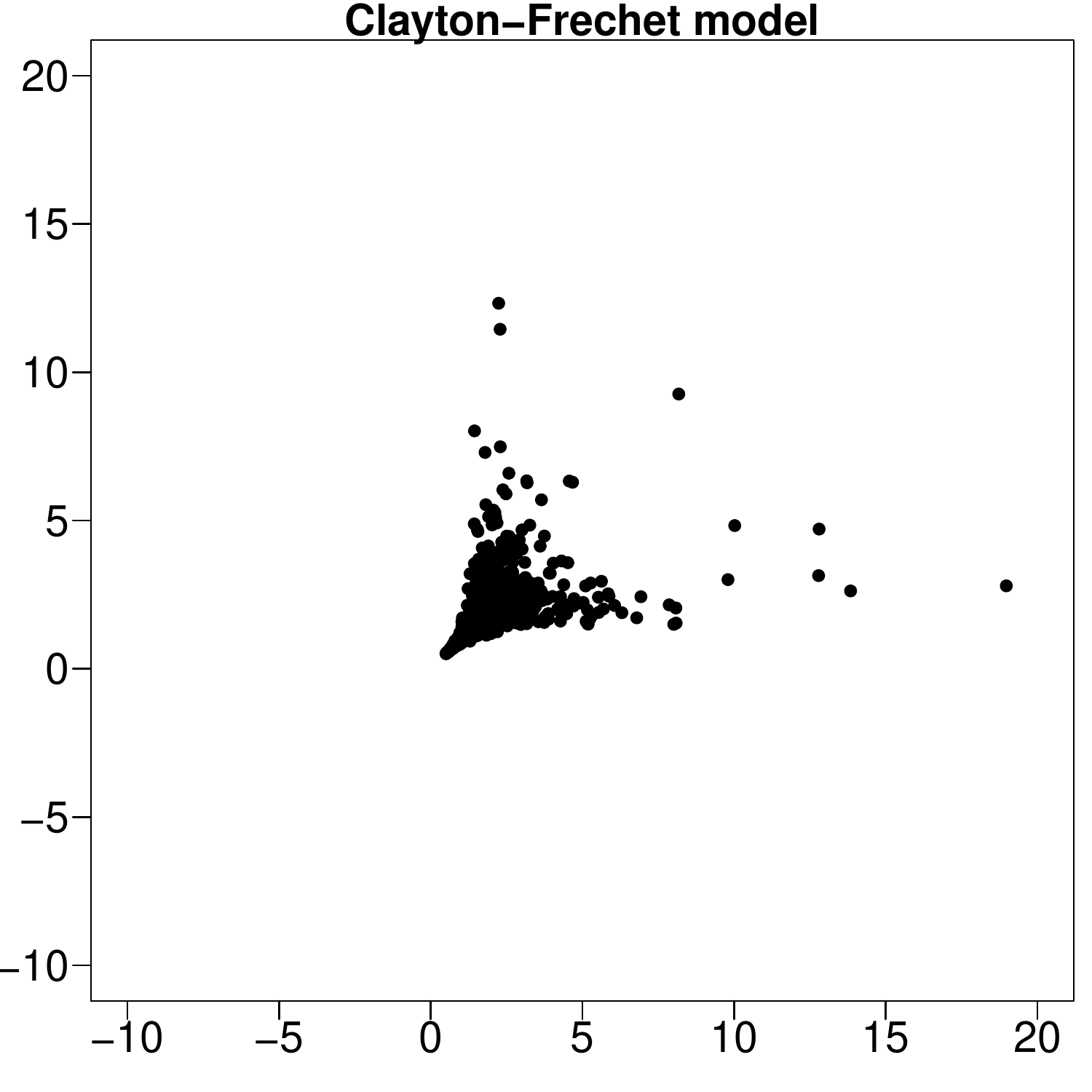}
\includegraphics[width=0.45\textwidth, page=2]{dependence_structures.pdf}\\
\includegraphics[width=0.45\textwidth, page=3]{dependence_structures.pdf}
\includegraphics[width=0.45\textwidth, page=4]{dependence_structures.pdf}
\caption{Examples of data samples generated from the models (i)-(iv) with sample size $n=2{,}500$ and dimension $d=2$.}
\label{fig:examples}
\end{figure}

\begin{landscape}
\begin{figure}
\centering
\includegraphics[width=0.32\textwidth, page=1]{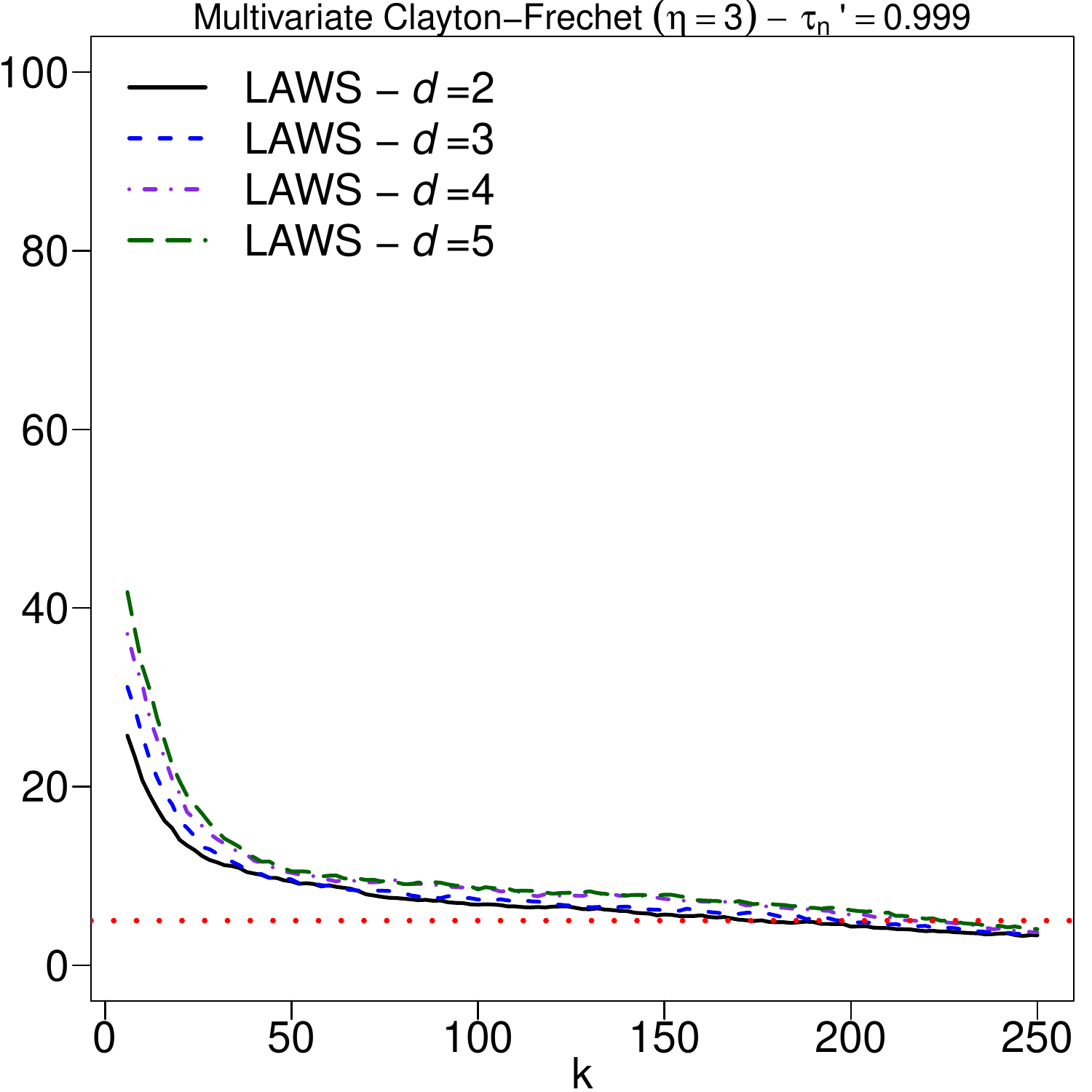}
\includegraphics[width=0.32\textwidth, page=3]{CovProbExtremeLevels0999.pdf}
\includegraphics[width=0.32\textwidth, page=5]{CovProbExtremeLevels0999.pdf}
\includegraphics[width=0.32\textwidth, page=7]{CovProbExtremeLevels0999.pdf}\\
\includegraphics[width=0.32\textwidth, page=2]{CovProbExtremeLevels0999.pdf}
\includegraphics[width=0.32\textwidth, page=4]{CovProbExtremeLevels0999.pdf}
\includegraphics[width=0.32\textwidth, page=6]{CovProbExtremeLevels0999.pdf}
\includegraphics[width=0.32\textwidth, page=8]{CovProbExtremeLevels0999.pdf}

\caption{Monte Carlo actual non-coverage probabilities (in \%) for the LAWS and QB confidence regions $\widetilde{\mathcal{E}}_{\tau'_n,\alpha}^{\star}$ and $\widehat{\mathcal{E}}_{\tau'_n,\alpha}^{\star}$, with $n=1{,}000$, $\tau_n'=0.999$ and $95\%$ nominal level, in models (i)-(iv). The horizontal dotted red line represents the $5\%$ nominal non-coverage probability.}
\label{fig:cov_prob_laws_QB_extreme}
\end{figure}
\end{landscape}

\begin{landscape}
\begin{figure}
\centering
\includegraphics[width=0.32\textwidth, page=1]{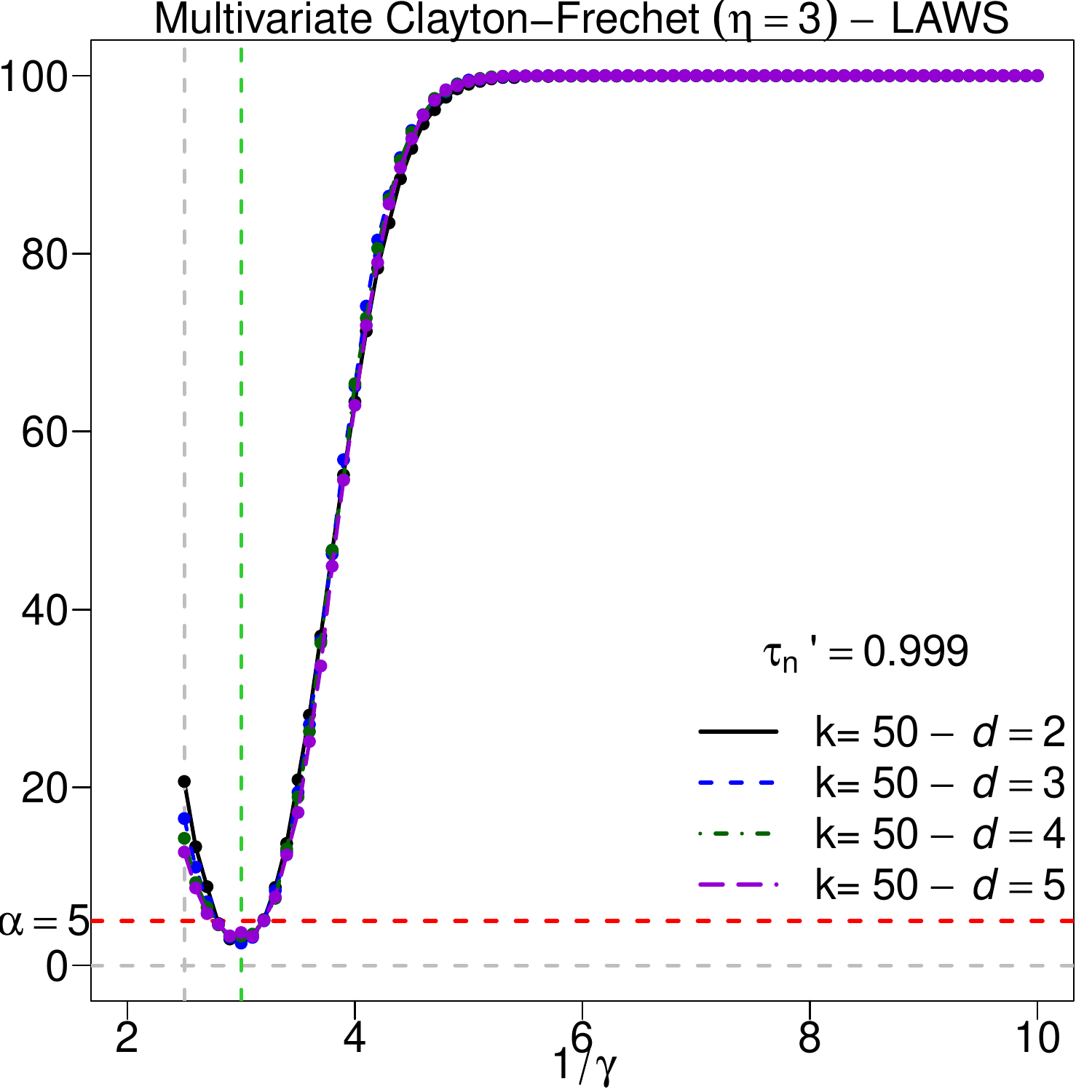}
\includegraphics[width=0.32\textwidth, page=3]{PowerTestAlldLAWSQB.pdf}
\includegraphics[width=0.32\textwidth, page=5]{PowerTestAlldLAWSQB.pdf}
\includegraphics[width=0.32\textwidth, page=7]{PowerTestAlldLAWSQB.pdf}\\
\includegraphics[width=0.32\textwidth, page=2]{PowerTestAlldLAWSQB.pdf}
\includegraphics[width=0.32\textwidth, page=4]{PowerTestAlldLAWSQB.pdf}
\includegraphics[width=0.32\textwidth, page=6]{PowerTestAlldLAWSQB.pdf}
\includegraphics[width=0.32\textwidth, page=8]{PowerTestAlldLAWSQB.pdf}
\caption{Monte Carlo rejection rate (in \%) of the LAWS and QB versions of the tests of equality of extreme expectiles, with $5\%$ nominal type I error rate (horizontal dotted red line), for $n=1{,}000$, $\tau_n'=0.999$ and $k=50$, in models (i)-(iv) where the first marginal is allowed to have a tail index varying in $[0.1,0.4]$.}
\label{fig:power_test}
\end{figure}
\end{landscape}

\begin{landscape}
\begin{figure}
\centering
\includegraphics[width=0.32\textwidth, page=1]{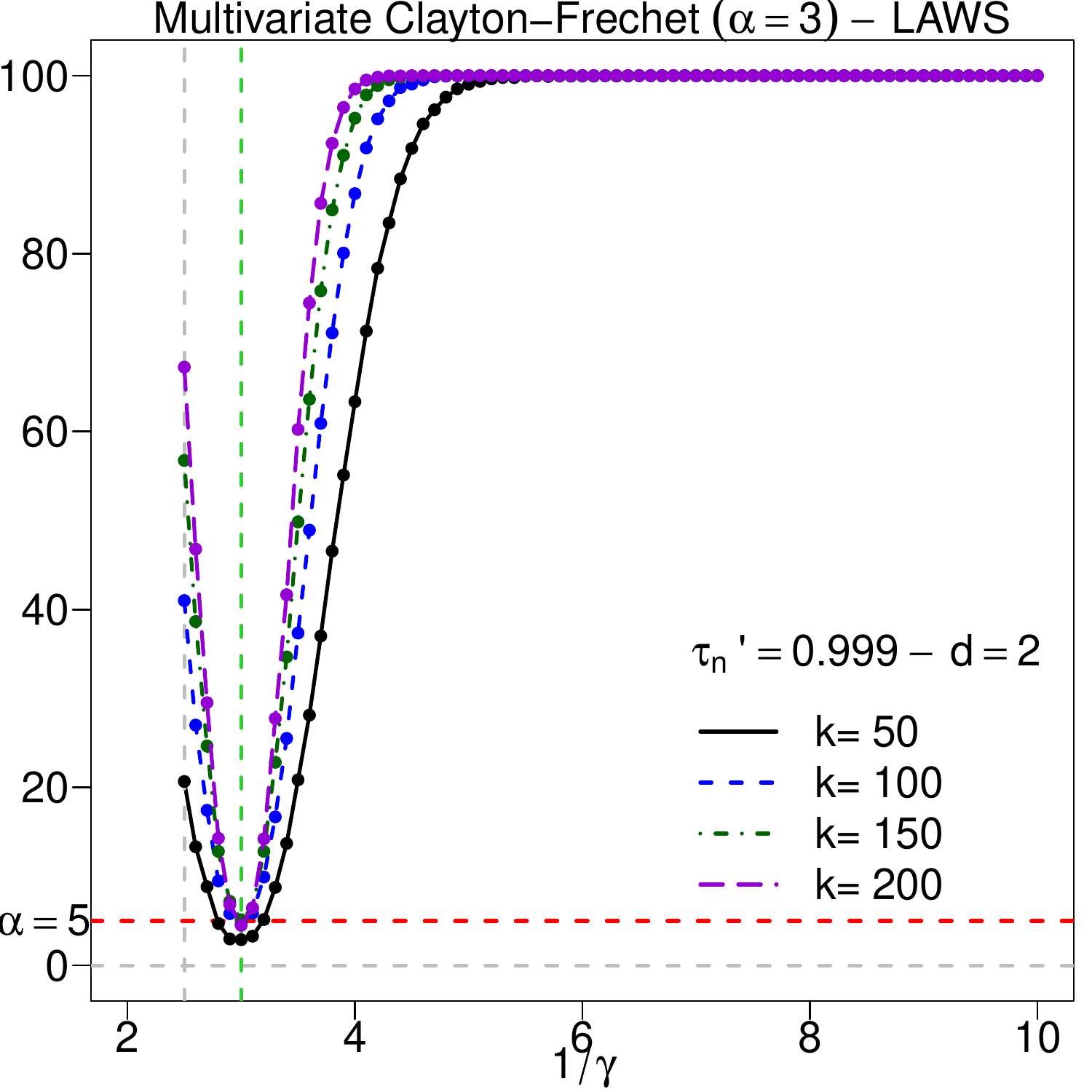}
\includegraphics[width=0.32\textwidth, page=3]{PowerTest2dAllModelsLAWSQB.pdf}
\includegraphics[width=0.32\textwidth, page=5]{PowerTest2dAllModelsLAWSQB.pdf}
\includegraphics[width=0.32\textwidth, page=7]{PowerTest2dAllModelsLAWSQB.pdf}\\
\includegraphics[width=0.32\textwidth, page=2]{PowerTest2dAllModelsLAWSQB.pdf}
\includegraphics[width=0.32\textwidth, page=4]{PowerTest2dAllModelsLAWSQB.pdf}
\includegraphics[width=0.32\textwidth, page=6]{PowerTest2dAllModelsLAWSQB.pdf}
\includegraphics[width=0.32\textwidth, page=8]{PowerTest2dAllModelsLAWSQB.pdf}

\caption{Monte Carlo rejection rate (in \%) of the LAWS and QB versions of the tests of equality of extreme expectiles, with $5\%$ nominal type I error rate (horizontal dotted red line), for $n=1{,}000$, $\tau_n'=0.999$ and dimension $d=2$, in models (i)-(iv) where the first marginal is allowed to have a tail index varying in $[0.1,0.4]$.}
\label{fig:cov_prob_laws_extreme}
\end{figure}
\end{landscape}

\begin{landscape}
\begin{figure}
\centering
\includegraphics[width=0.25\textwidth, page=1]{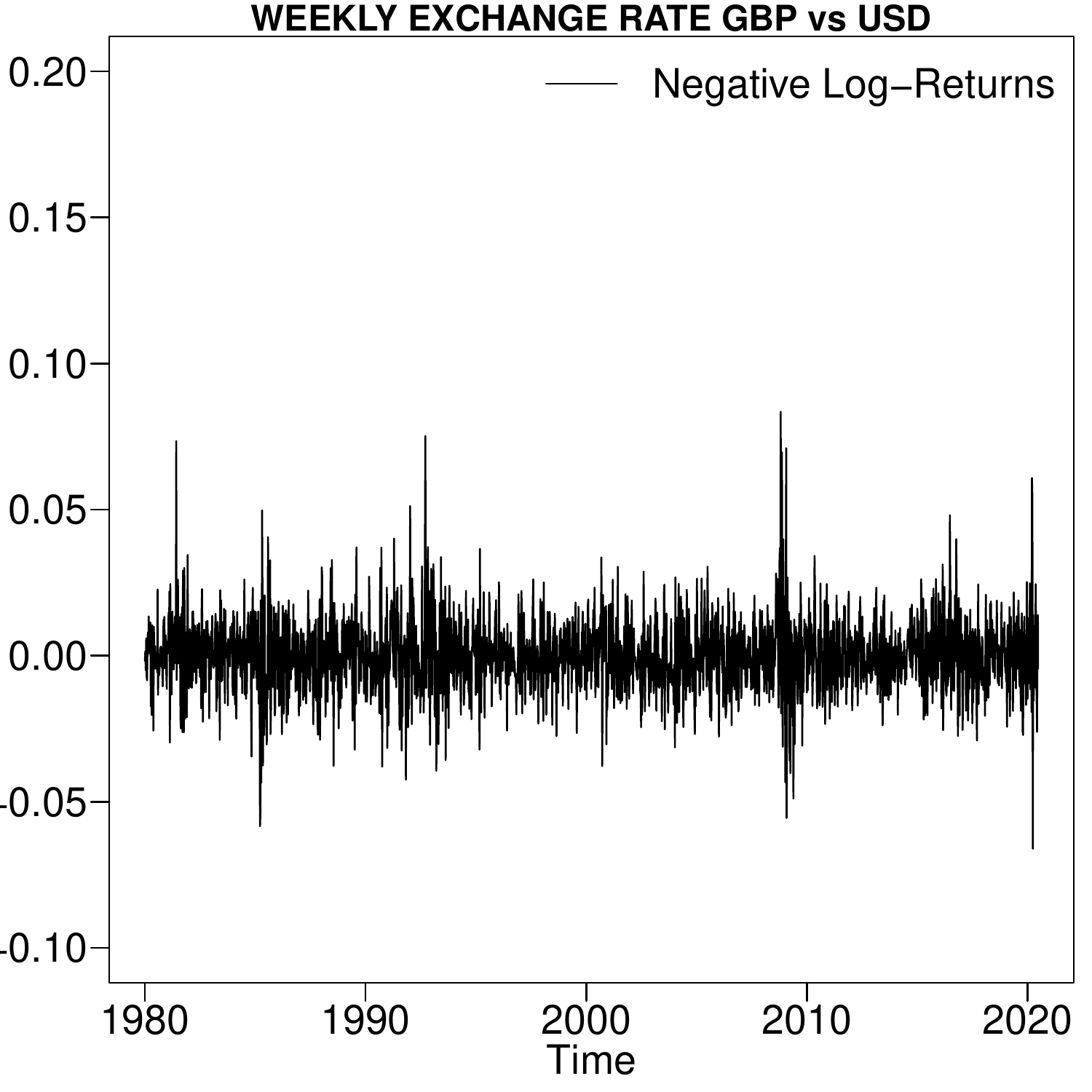}
\includegraphics[width=0.25\textwidth, page=1]{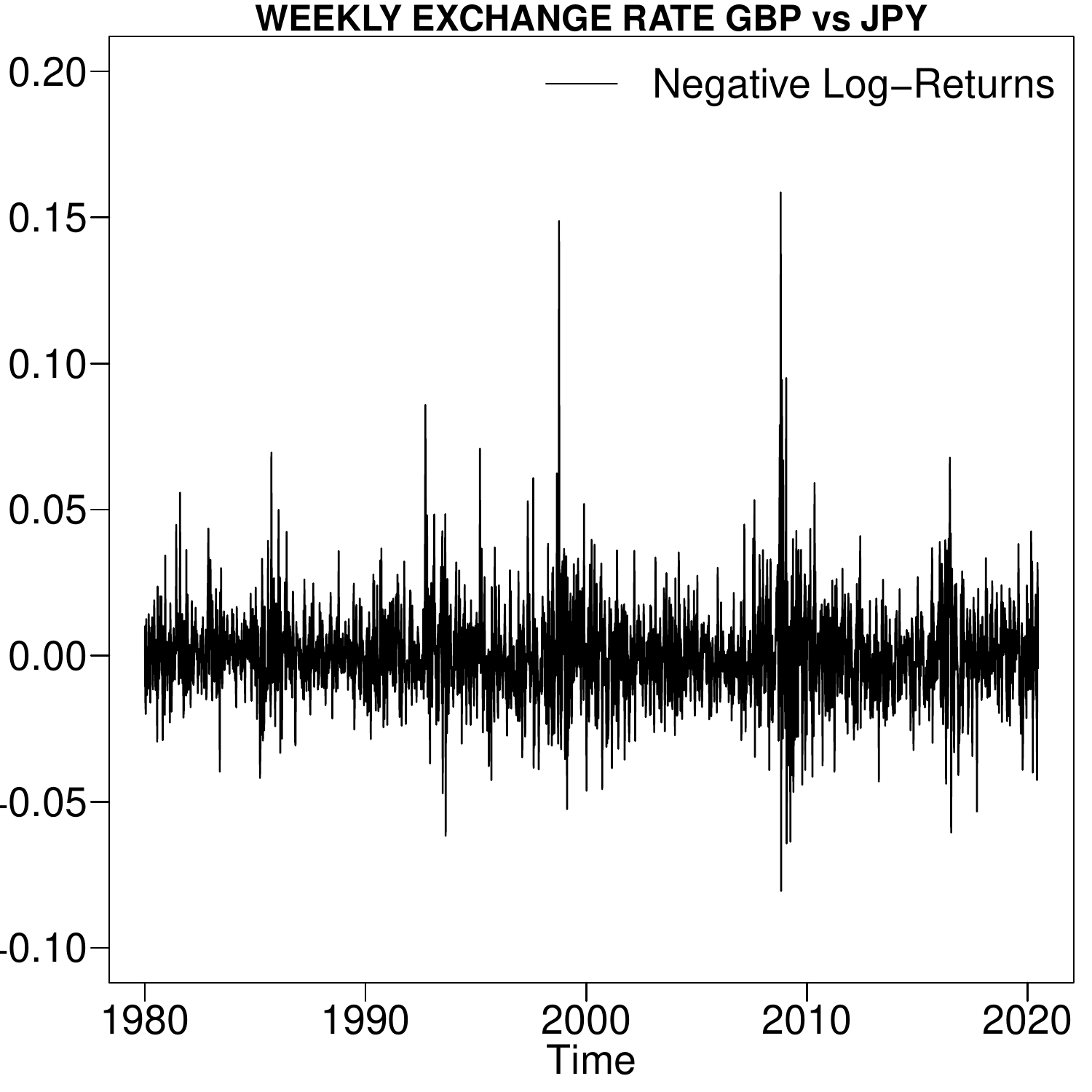}
\includegraphics[width=0.25\textwidth, page=1]{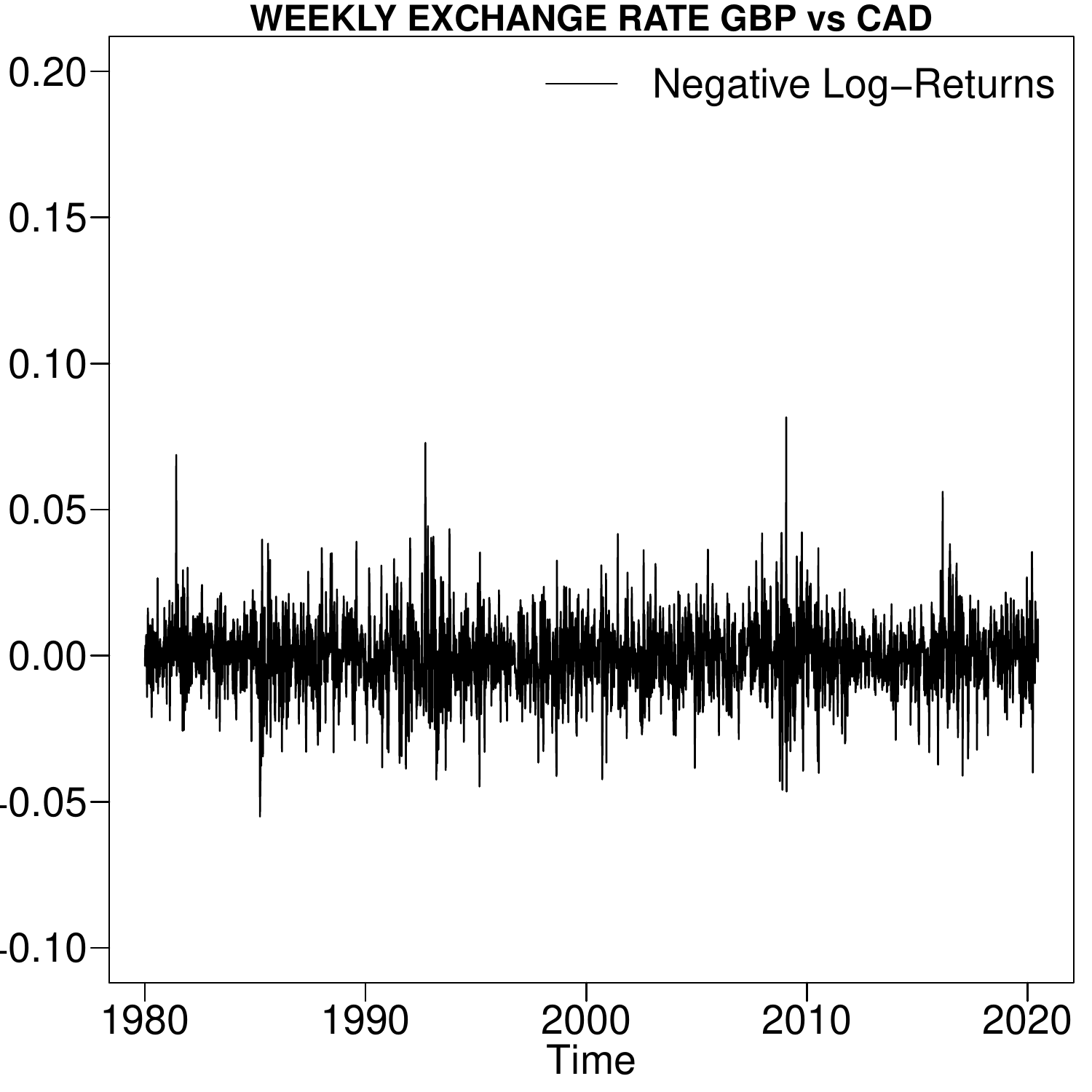}
\includegraphics[width=0.25\textwidth, page=1]{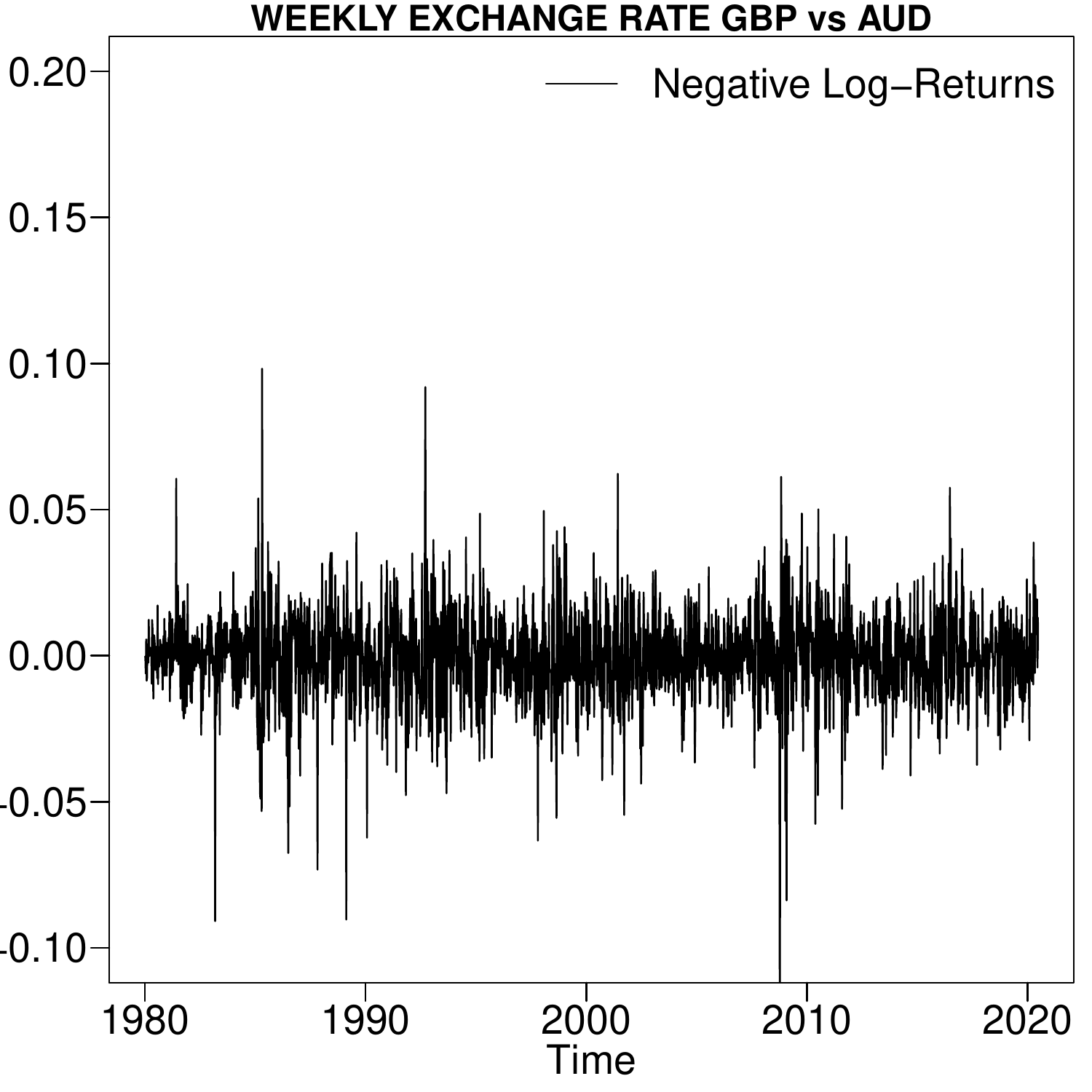}
\includegraphics[width=0.25\textwidth, page=1]{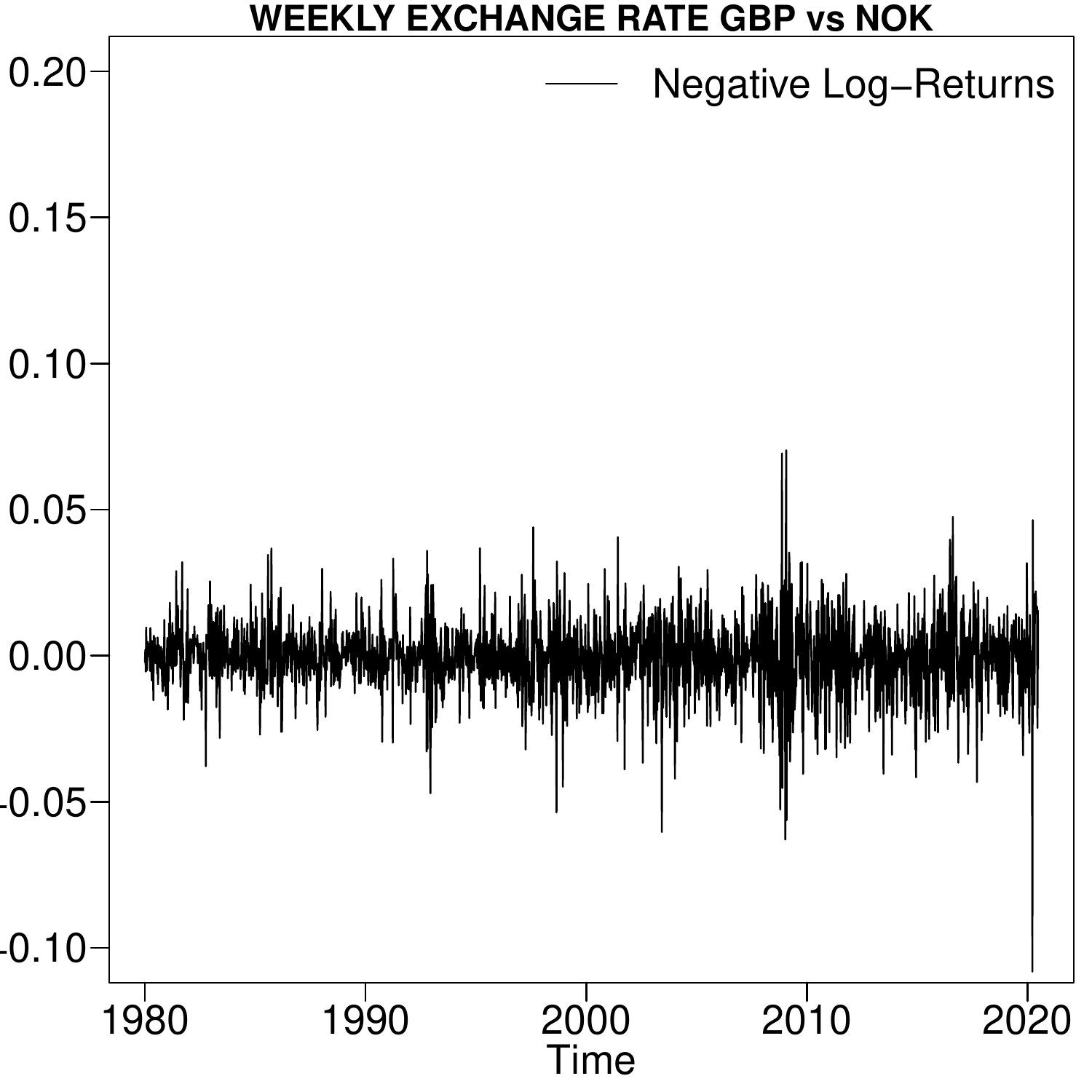}\\
\includegraphics[width=0.25\textwidth, page=5]{GBPUSD.pdf}
\includegraphics[width=0.25\textwidth, page=5]{GBPJPY.pdf}
\includegraphics[width=0.25\textwidth, page=5]{GBPCAD.pdf}
\includegraphics[width=0.25\textwidth, page=5]{GBPAUD.pdf}
\includegraphics[width=0.25\textwidth, page=5]{GBPNOK.pdf}\\
\includegraphics[width=0.25\textwidth, page=6]{GBPUSD.pdf}
\includegraphics[width=0.25\textwidth, page=6]{GBPJPY.pdf}
\includegraphics[width=0.25\textwidth, page=6]{GBPCAD.pdf}
\includegraphics[width=0.25\textwidth, page=6]{GBPAUD.pdf}
\includegraphics[width=0.25\textwidth, page=6]{GBPNOK.pdf}
\caption{Negative weekly log-returns of exchange rates (top row), tail index point estimates and asymptotic $95\%$ confidence intervals (Hill estimator, middle row) and expectile point estimates and asymptotic $95\%$ confidence intervals at the extreme level $\tau_n'=1-1/n=0.9995312$ obtained using the LAWS and QB extrapolating estimators (bottom row).}
\label{fig:ExcRatetailIndex}
\end{figure}
\end{landscape}

\begin{figure}[t!]
\centering
\includegraphics[width=0.7\textwidth]{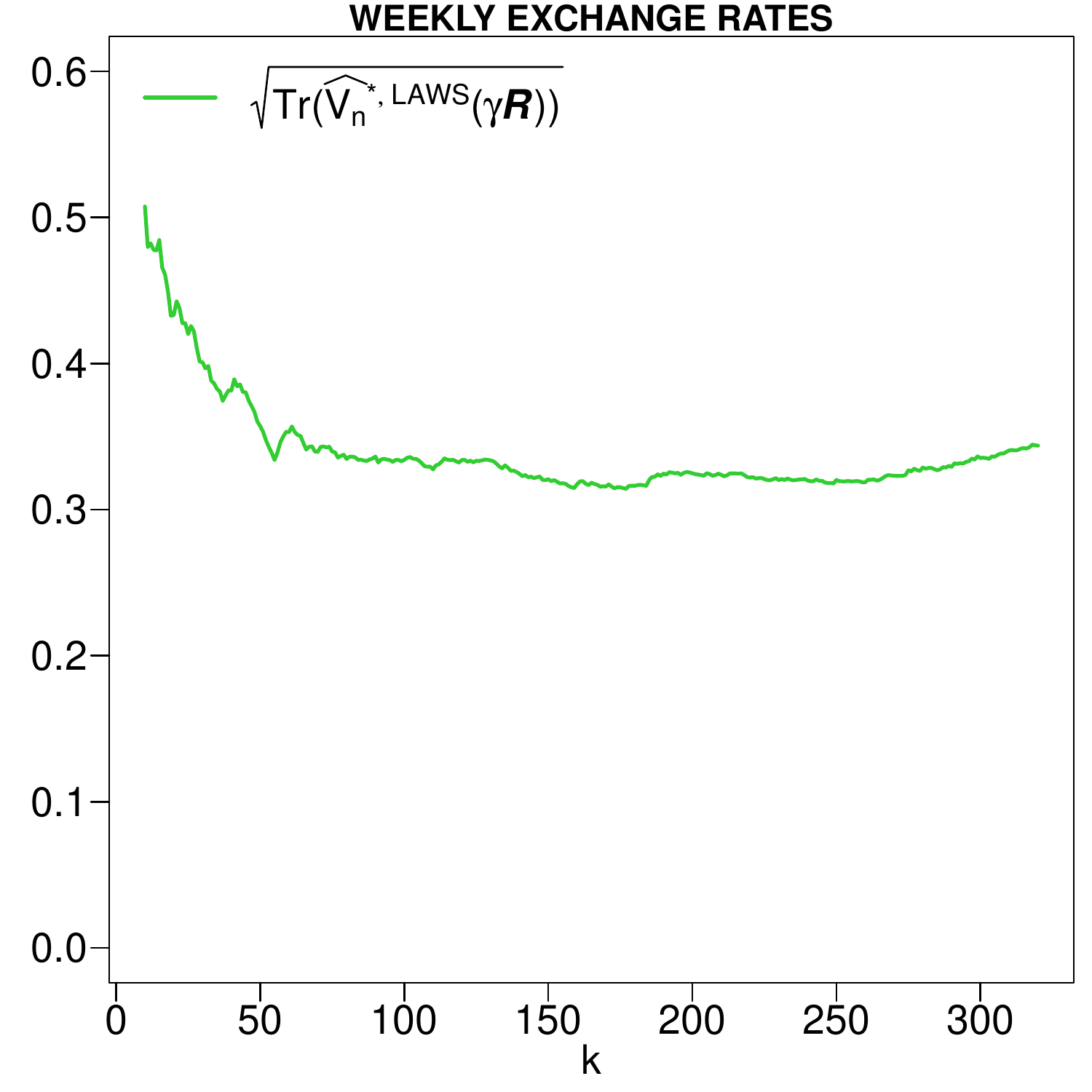}
\caption{Trace of $\protect\bfVhat_n^{\star,\mathrm{LAWS}}(\bfgamma,\bfR)$ as a function of $k$, obtained with $\tau_n'=0.9995312$.}
\label{fig:KSelectionTraceVar}
\end{figure}

\begin{figure}[t!]
\centering
\includegraphics[width=0.3\textwidth, page=4]{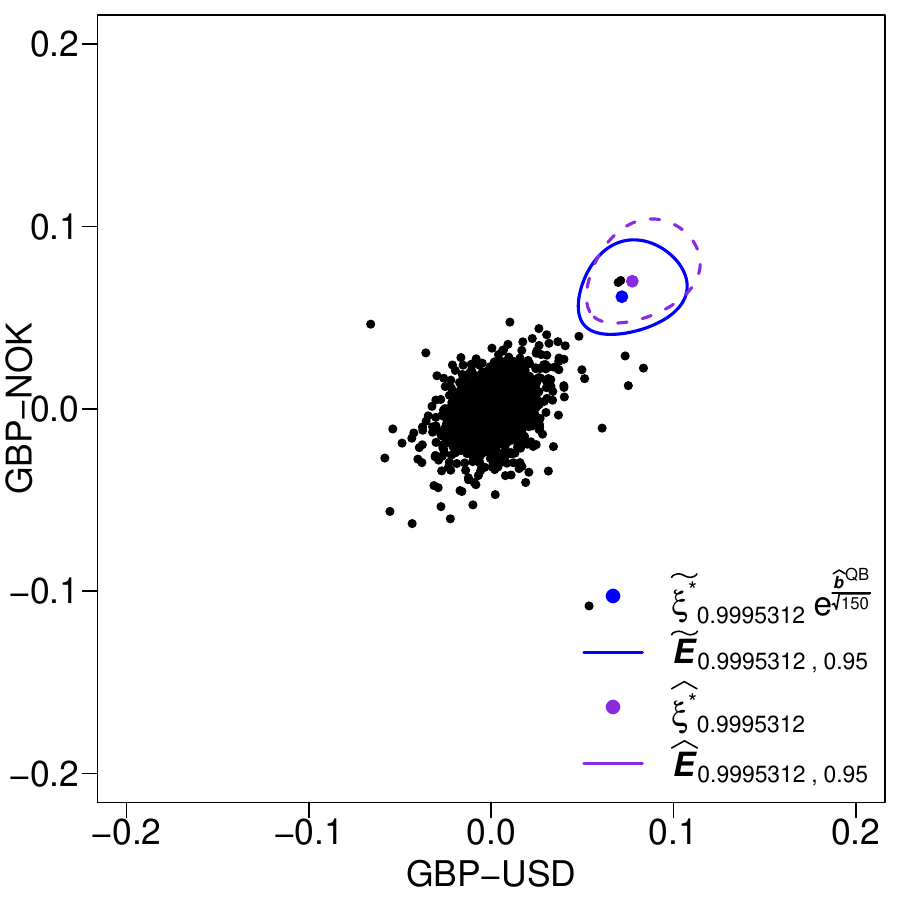}
\includegraphics[width=0.3\textwidth, page=5]{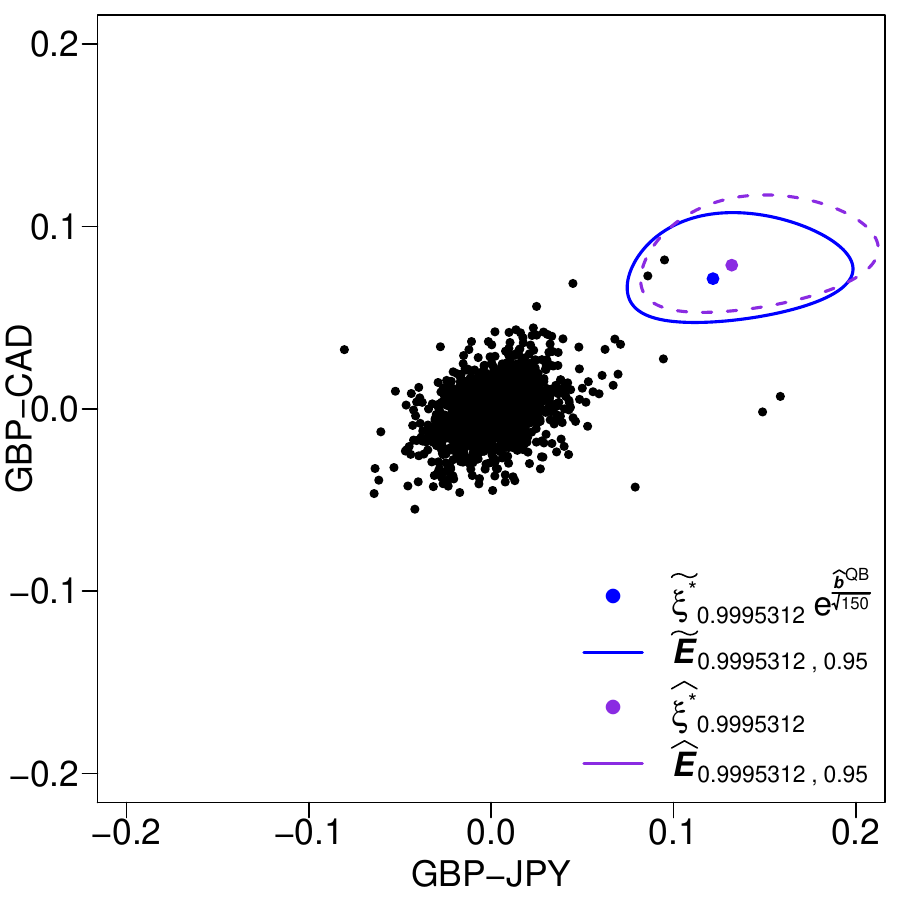}
\includegraphics[width=0.3\textwidth, page=6]{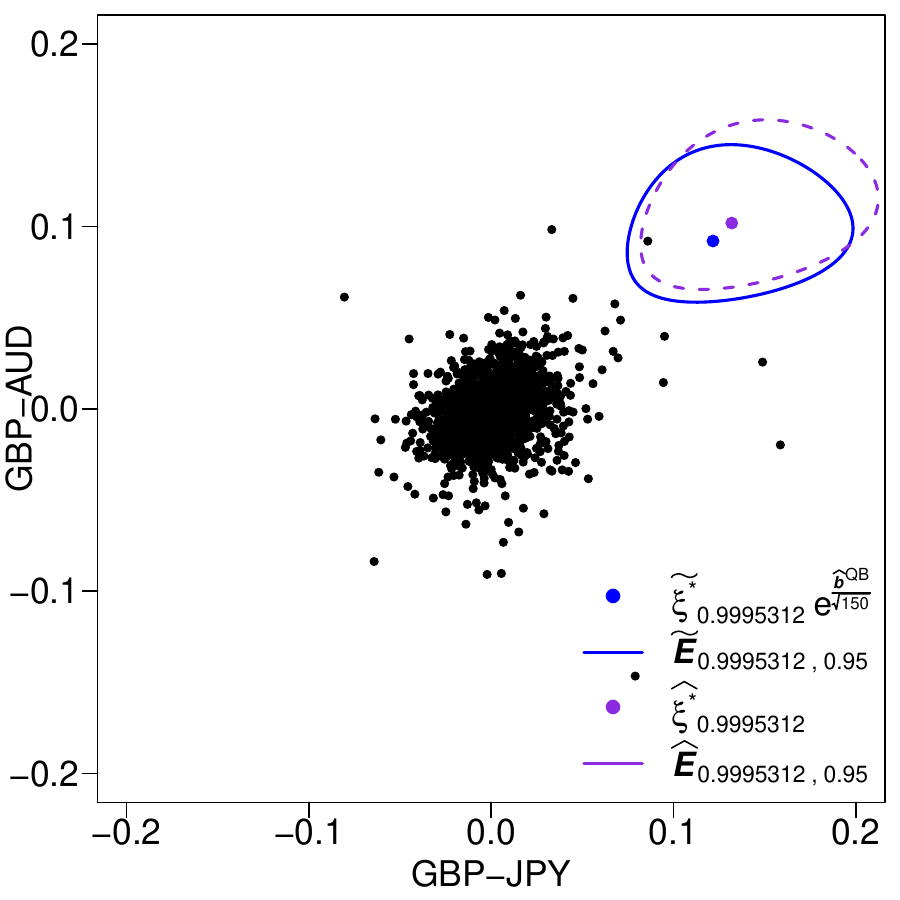}\\
\includegraphics[width=0.3\textwidth, page=7]{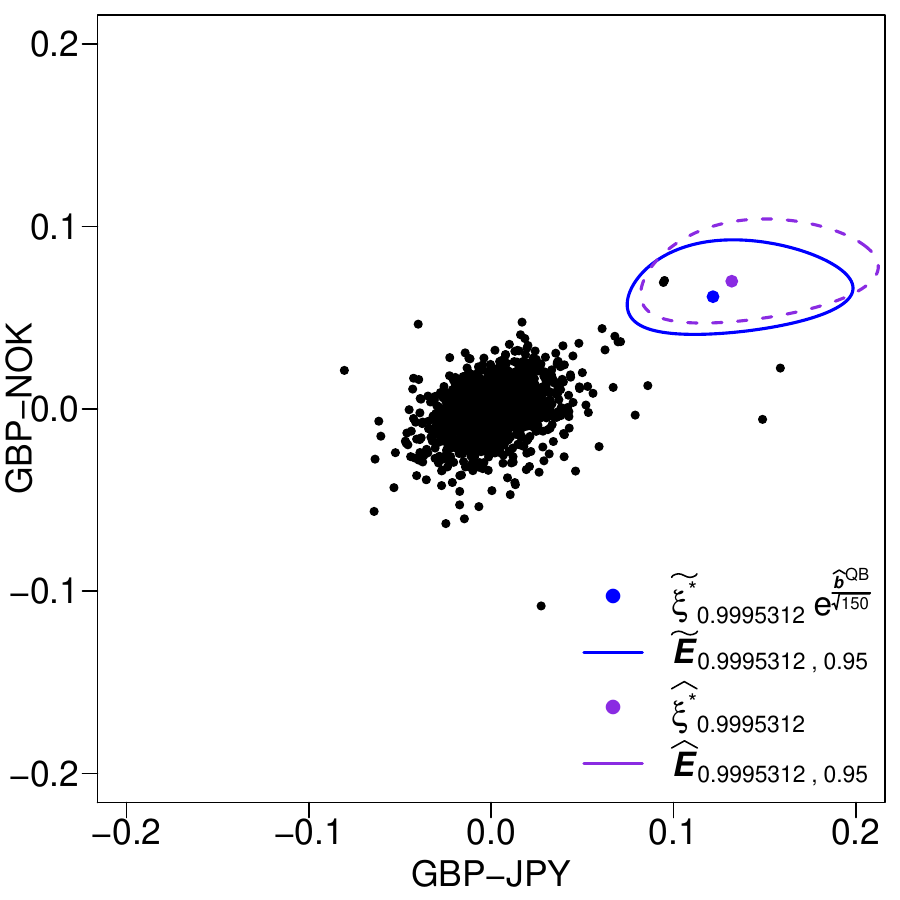}\\
\includegraphics[width=0.32\textwidth, page=5]{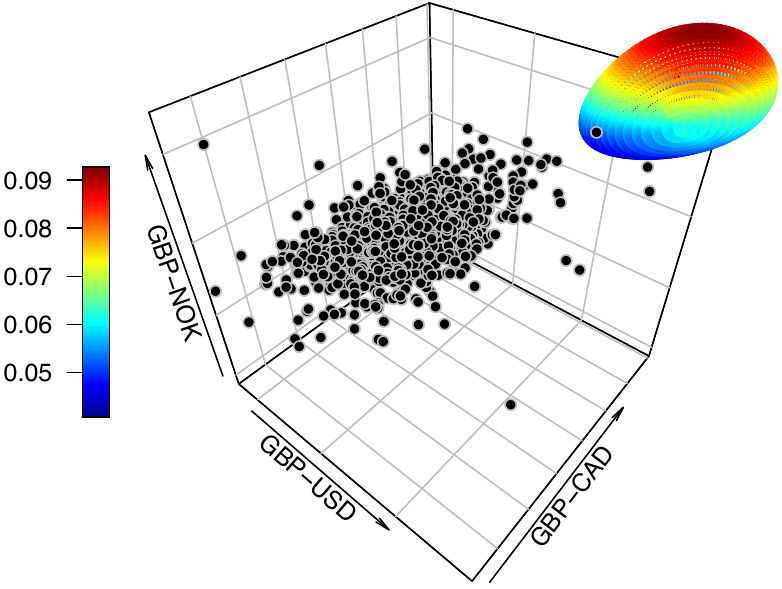}
\includegraphics[width=0.32\textwidth, page=7]{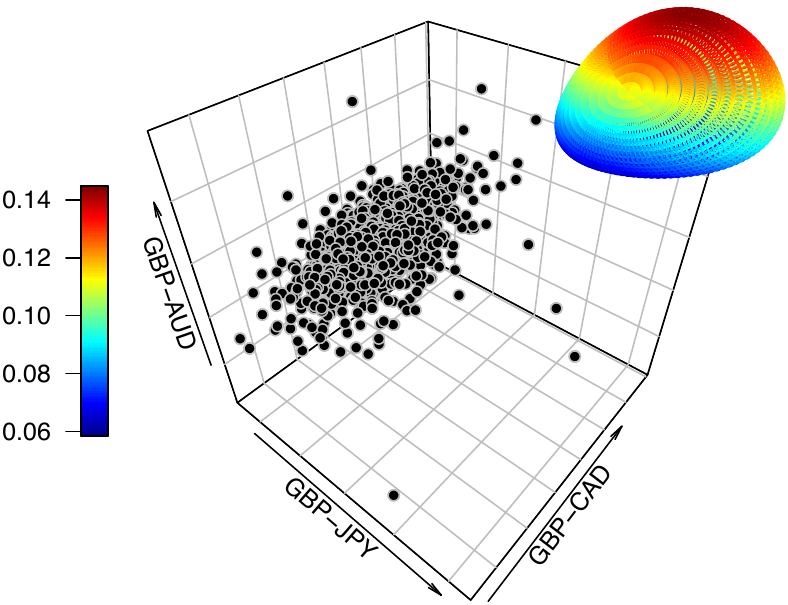}
\includegraphics[width=0.32\textwidth, page=8]{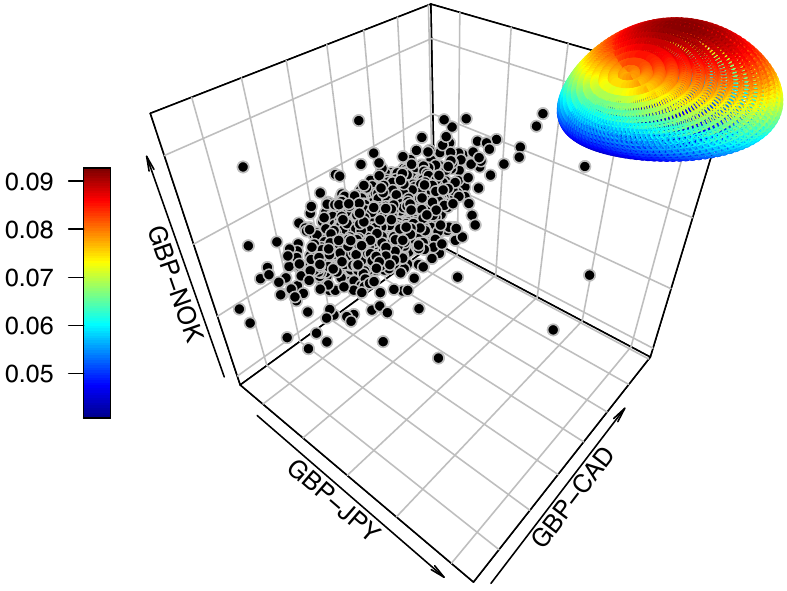}\\
\includegraphics[width=0.32\textwidth, page=9]{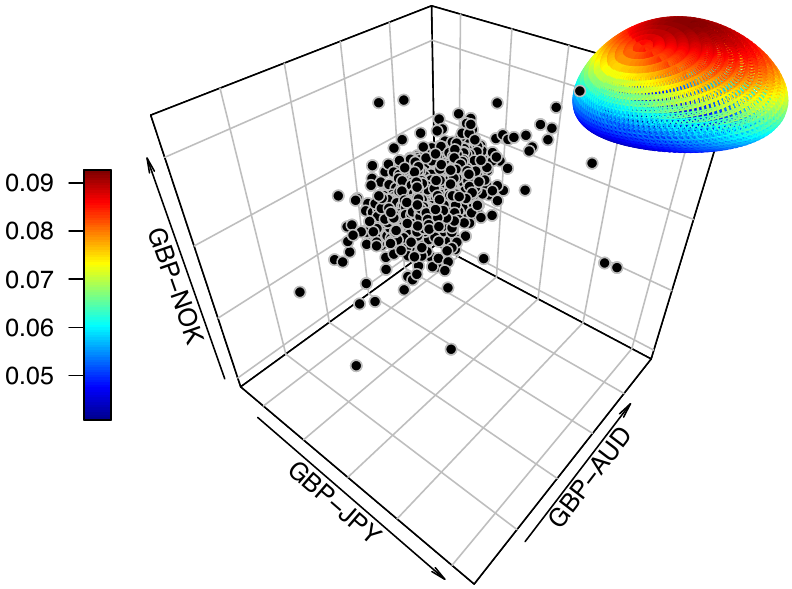}
\caption{Two- and three-dimensional $95\%$ confidence regions estimates for some pairs and triplets of exchange rate returns, obtained with $k=150$ and $\tau_n'=0.9995312$. In the three-dimensional case we only report LAWS-based confidence regions.}
\label{fig:PredictedConfidenceRegionsWeakDep}
\end{figure}

This section contains figures and tables linked to our results in Sections~\ref{sec:simulations} and~\ref{sec:real_data}. 
\begin{itemize}
\item Figure~\ref{fig:univariate_extreme_level} gives detailed results of our experiments concerning marginal uncertainty about tail expectiles, considered in Section~\ref{sec:simulations:1D}. 
\item Figures~\ref{fig:examples} 
and~\ref{fig:cov_prob_laws_QB_extreme}, as well as Tables~\ref{tab:RMSE_expect_LAWS_QB},~\ref{tab:cov_prob_expc_int_laws} and~\ref{tab:cov_prob_expc_int_QB} give further information about our experiments on joint tail expectile inference in Section~\ref{sec:simulations:joint}. 
\item Table~\ref{tab:significance_test} and Figures~\ref{fig:power_test} and~\ref{fig:cov_prob_laws_extreme} contain additional information related to our experiments on testing for equality of extreme expectiles in Section~\ref{sec:simulations:testing}. 
\item Figures~\ref{fig:ExcRatetailIndex},~\ref{fig:KSelectionTraceVar} and~\ref{fig:PredictedConfidenceRegionsWeakDep} in Section~\ref{sec:real_data} give additional results on tail index and extreme expectile estimates related to our real data analysis, as well as certain bivariate and trivariate confidence regions for extreme expectiles. 
\end{itemize}
%

%
%
%
%
%

%
%
%
%
%
%
%
%

%
%
%
\begin{table}[b!]
\centering
\begin{tabular}{lcccccccc}
\toprule
$m$ & \multicolumn{2}{c}{$d=2$} & \multicolumn{2}{c}{$d=3$} & 
\multicolumn{2}{c}{$d=4$} & \multicolumn{2}{c}{$d=5$}\\
& LAWS & QB & LAWS & QB & LAWS & QB & LAWS & QB\\
\midrule
  & \multicolumn{8}{c}{Clayton-Fr\'echet Model ($\eta=3$)}\\
\cdashlinelr{1-9}
$1.0$ & 9.236 & 17.119 & 9.652 & 17.049 & 10.423 & 17.044 &  9.708 &  17.102\\
$1.5$ & 8.689& 16.041 & 8.937 & 16.070 & 8.698 & 16.025 & 8.761 & 16.093\\
$2.0$ & 8.258 & 15.432 & 8.157 & 15.413 & 8.147 & 15.426 & 7.922 & 15.399\\
$2.5$ & 7.921 & 14.920 & 7.337 & 14.939 & 7.423 & 15.426 & 7.544 & 14.950\\
$5.0$ & 7.477 & 13.394 & 6.190 & 13.417 & 6.186 & 13.420 & 6.323 & 13.394\\
$10.0$ & 5.048 & 12.064 & 5.242 & 12.059 & 5.214 & 12.074 & 5.237 & 12.042\\
\midrule
  & \multicolumn{8}{c}{Gaussian-Student-$t$ Model ($\nu=3$)}\\
\cdashlinelr{1-9}
$1.0$ & 12.640 & 12.947 & 12.766 & 13.039 & 13.905 & 12.890 & 12.626 & 13.199\\
$1.5$ & 10.305 & 10.913 & 10.686 & 10.987 & 10.222 & 10.867 &  10.079 & 10.917\\
$2.0$ & 9.018 & 9.626 & 8.892 & 9.693 & 9.351 & 9.712 & 9.321 & 9.559\\
$2.5$ & 9.098 & 8.772 & 8.572 & 8.774 & 8.397 & 8.789 & 8.857 & 8.799\\
$5.0$ & 6.969 & 6.886 & 6.987 & 6.873 & 7.010 & 6.806 &  6.750 &  6.827\\
$10.0$ & 5.563 & 5.353 & 5.771 & 5.355 & 5.479 & 5.352 & 5.479 & 5.371\\
\midrule
 & \multicolumn{8}{c}{Gumbel-Fr\'echet Model ($\eta=3$)}\\
\cdashlinelr{1-9}
$1.0$ & 10.707 & 17.138 & 9.386 & 17.113 & 9.330 & 17.059 & 9.200 & 17.041\\
$1.5$ & 8.339 & 16.088 & 8.407 & 16.029 & 8.373 & 16.096 & 8.921 & 16.128\\
$2.0$ & 7.657 & 15.431 & 7.602 & 15.366 & 8.286 & 15.373 & 8.152 & 15.396\\
$2.5$ & 8.301 & 14.972 & 8.138 & 14.882 & 7.584 & 15.001 & 7.331 & 14.914\\
$5.0$ & 6.122 & 13.414 & 5.995 & 13.459 & 6.195 & 13.413 & 6.245 & 13.446\\
$10.0$ & 5.073 & 12.057 & 4.994 & 12.062 & 5.339 & 11.981 & 5.192 & 12.057\\
\midrule
  & \multicolumn{8}{c}{Multivariate Student-$t$ Model ($\nu=3$)}\\
\cdashlinelr{1-9}
$1.0$ & 11.361 & 12.906 & 11.986 & 13.040 & 11.762 & 12.988 & 12.204 & 13.097\\
$1.5$ & 10.079 & 10.755 & 9.899 & 10.754 & 9.966 & 10.765 & 10.497 & 10.826\\
$2.0$ & 9.518 & 9.799 & 9.559 & 9.632 & 9.165 & 9.516 & 9.271 & 9.580\\
$2.5$ & 10.458 & 8.858 & 9.435 & 8.823 & 8.577 & 8.930 &  9.627 &  8.888\\
$5.0$ & 6.737 & 6.791 & 6.857 & 6.777 & 7.202 & 6.837 & 7.732 & 6.782\\
$10.0$ & 5.513 & 5.381 & 5.605 & 5.407 & 5.472 & 5.348 & 5.508 & 5.344\\
\bottomrule
\end{tabular}
\caption{Monte Carlo relative MSE (averaged across marginals and reported in $\sqrt{\cdot}\times 100$) of the LAWS and QB intermediate estimators at level $\tau_n$, with $n=m\cdot 10^3$ (left column), $\tau_n = 1-1/\sqrt{n}$ and $95\%$ nominal level.}
\label{tab:RMSE_expect_LAWS_QB}
\end{table}
%
%
%
%

%
%
%
\begin{table}[htbp]
\centering
\begin{tabular}{lcccc}
\toprule
$m$ & $d=2$ & $d=3$ & $d=4$ & $d=5$\\
  & \multicolumn{4}{c}{Clayton-Fr\'echet Model ($\eta=3$)}\\
\cdashlinelr{1-5}
$1.0$ & 5.16(11.72) & 5.42(14.08) & 5.72(15.87) &  5.81(17.65) \\
$1.5$ & 4.88(10.76) & 5.05(12.52) & 5.06(13.85) & 4.96(14.97) \\
$2.0$ & 4.85(10.53) & 4.60(11.81) & 5.24(13.72) & 5.09(14.57) \\
$2.5$ & 4.83(10.36) & 4.71(11.15) & 4.93(12.63) & 4.73(14.53) \\
$5.0$ & 4.38(~8.71) & 4.90(10.18) & 4.72(11.84) & 4.41(11.70) \\
$10.0$ & 4.20(~8.07) & 4.25(~9.26) & 4.40(10.17) & 4.31(10.92) \\
\midrule
  & \multicolumn{4}{c}{Gaussian-Student-$t$ Model ($\nu=3$)}\\
\cdashlinelr{1-5}
$1.0$ & 5.49(~8.37) & 6.08(~9.53) & 6.61(10.49) & 6.38(11.23) \\
$1.5$ & 5.28(~7.39) & 5.86(~8.61) & 6.29(~9.59) & 6.34(10.20) \\
$2.0$ & 5.31(~7.55) & 5.32(~7.93) & 5.43(~8.24) & 5.92(~9.43) \\
$2.5$ &  5.08(~6.95) & 5.22(~7.47) & 5.80(~7.88) & 6.08(~8.96) \\
$5.0$ & 4.78(~6.44) & 5.15(~7.11) & 5.48(~7.67) & 5.41(~7.52)\\
$10.0$ & 4.94(~6.51) & 5.49(~6.79) & 5.85(~7.25) & 5.04(~7.01)\\
\midrule
  & \multicolumn{4}{c}{Gumbel-Fr\'echet Model ($\eta=3$)}\\
\cdashlinelr{1-5}
$1.0$ & 5.13(12.09) & 4.65(12.98) & 4.84(15.16) & 4.10(15.35) \\
$1.5$ & 4.28(10.77) & 4.10(12.19) & 3.99(13.94) & 3.70(15.16) \\
$2.0$ & 4.12(10.59) & 4.18(12.59)& 3.75(13.77) & 3.27(14.47) \\
$2.5$ & 3.97(10.60) & 3.76(12.33) & 3.60(14.15) & 3.31(14.62) \\
$5.0$ & 3.72(~9.91) & 3.34(11.65) & 3.06(14.00) & 3.11(14.26) \\
$10.0$ & 3.34(~9.45) & 2.91(11.45) & 2.57(12.09) & 2.52(14.50) \\
\midrule
  & \multicolumn{4}{c}{Multivariate Student-$t$ Model ($\nu=3$)}\\
\cdashlinelr{1-5}
$1.0$ & 4.76(~8.33) & 5.84(10.29) & 6.08(10.49) & 6.22(11.90) \\
$1.5$ & 4.89(~8.06) & 5.27(~9.08) & 5.87(~9.84) & 5.49(10.82) \\
$2.0$ & 4.49(~7.42) & 4.85(~8.04) & 5.33(~9.40) & 5.53(10.19) \\
$2.5$ & 4.71(~7.61) & 5.46(~9.09) & 5.27(~9.47) & 5.03(~9.75) \\
$5.0$ & 4.31(~7.19) & 4.66(~7.86) & 6.10(~8.21) & 5.02(~9.50) \\
$10.0$ & 3.89(~6.37) & 4.34(~7.42) & 5.19(~8.61) & 4.26(~9.15) \\
\bottomrule
\end{tabular}
\caption{Monte Carlo actual non-coverage probability (in \%) for the LAWS confidence region estimator $\widetilde{\mathcal{E}}_{\tau_n,\alpha}$ at the intermediate level, with $n=m\cdot 10^3$ (left column), $\tau_n=1-1/\sqrt{n}$ and $95\%$ nominal level. Between brackets we report the coverage probability obtained assuming independence between the margins.}
\label{tab:cov_prob_expc_int_laws}
\end{table}
%
%
%

%
%
%
%
\begin{table}[t!]
\centering
\begin{tabular}{lcccc}
\toprule
$m$ & $d=2$ & $d=3$ & $d=4$ & $d=5$\\
  & \multicolumn{4}{c}{Clayton-Fr\'echet Model ($\alpha=3$)}\\
\cdashlinelr{1-5}
$1.0$ & 6.52(0.01) & 7.10(0.02) & 7.65(0.00) & 8.82(0.01) \\
$1.5$ & 5.72(0.00) & 6.46(0.01) & 7.32(0.00) & 7.43(0.00) \\
$2.0$ & 5.58(0.00) & 6.05(0.00) & 6.66(0.00) & 7.45(0.00) \\
$2.5$ & 5.49(0.00) & 6.11(0.00) & 6.71(0.00) & 6.93(0.00) \\
$5.0$ & 4.73(0.00) & 5.11(0.00) & 5.88(0.00) & 5.79(0.00) \\
$10.0$ & 4.64(0.00) & 5.16(0.00) & 5.26(0.00) & 5.07(0.00) \\
\midrule
  & \multicolumn{4}{c}{Gaussian-Student-$t$ Model ($\nu=3$)}\\
\cdashlinelr{1-5}
$1.0$ & 6.42(0.13) & 6.26(0.09) & 7.00(0.04) & 8.16(0.04) \\
$1.5$ & 5.83(0.01) & 6.63(0.04) & 6.78(0.01) & 6.76(0.00) \\
$2.0$ & 5.55(0.04) & 6.04(0.01) & 6.66(0.00) & 6.67(0.00) \\
$2.5$ & 5.48(0.00) & 6.15(0.00) & 6.30(0.00) & 6.30(0.00) \\
$5.0$ & 5.44(0.00) & 5.96(0.00) & 6.15(0.00) & 6.08(0.00)\\
$10.0$ & 5.43(0.00) & 5.75(0.00) & 5.37(0.00) & 6.22(0.00)\\
\midrule
  & \multicolumn{4}{c}{Gumbel-Fr\'echet Model ($\alpha=3$)}\\
\cdashlinelr{1-5}
$1.0$ & 4.54(0.21) & 4.07(0.29) & 3.93(0.35) & 3.46(0.36) \\
$1.5$ & 3.77(0.06) & 3.52(0.06) & 3.33(0.15) & 3.39(0.23) \\
$2.0$ & 4.03(0.07) & 3.65(0.05) & 3.11(0.12) & 3.07(0.07) \\
$2.5$ & 3.69(0.00) & 3.55(0.04) & 3.56(0.09) & 2.82(0.12) \\
$5.0$ & 3.63(0.00) & 3.23(0.01) & 2.83(0.03) & 2.83(0.03) \\
$10.0$ & 3.11(0.00) & 2.81(0.00) & 2.68(0.00) & 2.69(0.00) \\
\midrule
  & \multicolumn{4}{c}{Multivariate Student-$t$ Model ($\nu=3$)}\\
\cdashlinelr{1-5}
$1.0$ & 5.23(0.13) & 5.76(0.05) & 6.30(0.05) & 6.87(0.04) \\
$1.5$ & 4.91(0.03) & 5.47(0.06) & 5.37(0.04) & 6.08(0.01) \\
$2.0$ & 5.21(0.03) & 5.61(0.02) & 5.88(0.02) & 5.20(0.00) \\
$2.5$ & 5.23(0.02) & 5.25(0.00) & 5.68(0.00) & 5.58(0.00) \\
$5.0$ & 4.55(0.00) & 4.85(0.00) & 4.97(0.00) & 5.18(0.00) \\
$10.0$ & 4.77(0.00) & 4.56(0.00) & 5.02(0.00) & 5.15(0.00) \\
\bottomrule
\end{tabular}
\caption{Monte Carlo actual non-coverage probability (in \%) for the QB confidence region estimator $\widehat{\mathcal{E}}_{\tau_n,\alpha}$ at the intermediate level, with $n=m\cdot 10^3$ (left column), $\tau_n=1-1/\sqrt{n}$ and $95\%$ nominal level. Between brackets we report the coverage probability obtained assuming independence between the margins.}
\label{tab:cov_prob_expc_int_QB}
\end{table}
%
%
%
%
%

%
%
%
%
%
%
%
%
%
%

%
%
%

%

%
%
%
%
\begin{table}[t!]
\centering
\begin{tabular}{cccccc}
\toprule
Model & Method & $d=2$ & $d=3$ & $d=4$ & $d=5$\\
\midrule
(i) & LAWS & ~2.89 & ~3.28 & ~3.51 & ~3.68 \\
 & QB & ~8.71 & ~8.87 & 10.70 & 10.72 \\
\cdashlinelr{1-6}
(ii) & LAWS & ~2.93 & ~3.60 &  ~4.62 & ~5.12 \\
 & QB & ~4.57 & ~3.54 & ~3.36 & ~3.21 \\
\cdashlinelr{1-6}
(iii) & LAWS & ~2.26 & ~2.25 &  ~2.21 & ~2.06 \\
 & QB & ~2.79 & ~2.85 & ~1.96 & ~1.57 \\
\cdashlinelr{1-6}
(iv) & LAWS & ~2.44 & ~3.43 &  ~3.97 & ~4.26 \\
 & QB & ~4.06 & ~3.10 & ~2.77 & ~2.48 \\
\bottomrule
\end{tabular}
\caption{Monte Carlo rejection rate (in \%) of the tests of equality of extreme expectiles, with $5\%$ nominal type I error rate, for $n=1{,}000$, $\tau_n'=0.999$ and $k=50$.}
\label{tab:significance_test}
\end{table}
%
%
%

%

%

%
%
%
%

%
%
%
%
%
%
%
%
%
%
%
%
%
%
%
%
%

%
\end{document}